\documentclass[%
 reprint,
superscriptaddress,
nofootinbib,
 amsmath,amssymb,
 aps,
 prd,
floatfix,
]{revtex4-2}
\usepackage{graphicx}
\usepackage{bm}
\usepackage{epsfig}
\usepackage{amsmath}
\usepackage{natbib}
\usepackage{hyperref}
\usepackage{color}

\begin{document}

\title{Don't cross the streams: caustics from Fuzzy Dark Matter}
\author{Neal Dalal}
\email{ndalal@pitp.ca}
\affiliation{Perimeter Institute for Theoretical Physics, 31 Caroline Street N., Waterloo, Ontario, N2L 2Y5, Canada}
\author{Jo Bovy}
\affiliation{David A. Dunlap Department of Astronomy and Astrophysics, University of Toronto, 50 St. George Street, Toronto, ON, M5S 3H4, Canada}
\author{Lam Hui}
\affiliation{
Center for Theoretical Physics, Department of Physics, Columbia University, New York, NY 10027, USA
}
\author{Xinyu Li}
\affiliation{Perimeter Institute for Theoretical Physics, 31 Caroline Street N., Waterloo, Ontario, N2L 2Y5, Canada}
\affiliation{%
Canadian Institute for Theoretical Astrophysics, 60 St.\ George St., Toronto, ON M5R 2M8, Canada}

\begin{abstract}
We study how tidal streams from globular clusters may be used to constrain the mass of ultra-light dark matter particles, called `fuzzy' dark matter (FDM).  A general feature of FDM models is the presence of ubiquitous density fluctuations in bound, virialized dark matter structures, on the scale of the de Broglie wavelength, arising from wave interference in the evolving dark matter distribution.  These time-varying fluctuations can disturb the motions of stars, leading to potentially observable signatures in cold thin tidal streams in our own Galaxy.  The study of this effect has been hindered by the difficulty in simulating the FDM wavefunction in Milky Way-sized systems.  We present a simple method to evolve realistic wavefunctions in nearly static potentials, that should provide an accurate estimate of this granulation effect.  We quantify the impact of FDM perturbations on tidal streams, and show that initially, while stream perturbations are small in amplitude, their power spectra exhibit a sharp cutoff corresponding to the de Broglie wavelength of the FDM potential fluctuations.  Eventually, when stream perturbations become nonlinear, 
fold caustics generically arise that lead to density fluctuations with universal behavior.  This erases the signature of the de Broglie wavelength in the stream density power spectrum, but we show that it will still be possible to determine the FDM mass in this regime, by considering the fluctuations in quantities like angular momenta or actions.
\end{abstract}

\maketitle

\section{Introduction}
\label{sec:intro}

The nature of dark matter remains an important outstanding problem in cosmology.
In recent years, the Fuzzy Dark Matter (FDM) model \citep{Hu:2000ke} has been proposed as a physically well-motivated variant of dark matter with a number of attractive properties \citep{Hui:2016ltb}.  In this model, dark matter is hypothesized to be an axion-like particle with an ultra-light mass, $m \gtrsim 10^{-22}$ eV, so small that the de Broglie wavelength of DM particles in our Galaxy is astronomical in scale, $\lambda = h/(mv) \lesssim 600$ pc.  The occupation number of DM particles in FDM models is so large that we can describe the dark matter using a classical, coherent wavefunction that obeys the coupled Schr\"odinger and Poisson equations \cite{Widrow:1993qq}. 
Detailed numerical simulations of cosmological structure formation have shown that FDM models can reproduce the spectacular successes of standard cold dark matter (CDM) on scales larger than galaxies, while deviating from CDM in interesting ways on sub-galactic scales \citep[e.g.,][]{Schive:2014dra,Schive:2014hza,Mocz:2015sda,Veltmaat:2016rxo,Schwabe:2016rze,Zhang:2016uiy,Mocz:2017wlg,Zhang:2017chj,Irsic:2017yje,Li:2018kyk}.  

For example, one unique aspect of FDM models with ultra-light masses is the ubiquitous presence of density fluctuations with ${\cal O}(1)$ density contrast on length scales of order the de Broglie wavelength, throughout bound virialized structures like dark matter halos \cite{Schive:2014dra,Hui:2020hbq}.  These granular fluctuations occur because of the interference between bound waves in halos, and are therefore generic in FDM models, independent of any additional physics like self-interactions.  These interference fringes can have potentially detectable effects on the dynamics of observed stars in galaxies, by producing gravitational perturbations.  \citet{Hui:2016ltb} provide a simple, order-of-magnitude estimate of the gravitational perturbations produced by FDM granularity, which we can summarize as follows (for a different toy model estimate of this effect, see also \cite{Amorisco2018}).

Let us approximate the interference fringes as density fluctuations of amplitude $\delta\rho\sim\rho$ and size $\delta r \sim \lambda/2\pi$, where $\rho$ is the mean local density of dark matter and $\lambda$ is the de Broglie wavelength.  The corresponding potential fluctuation is $\delta\Phi \equiv \sigma_v^2 \propto \delta\rho \, \delta r^2$.  A test particle passing through this potential fluctuation at velocity $v$ will be deflected by an angle $(\sigma_v/v)^2$, giving a transverse velocity kick of $\delta v \approx \sigma_v^2/v$. Test particles therefore experience random walks in their transverse velocities, where each step in the random walk has size $\delta v$, and the number of steps is $N = v\,t/\delta r$, where $t$ is the duration of time over which the test particle moves through FDM perturbations.  If we assume that $v=200\,$km/s, $t=5\,$ Gyr, $\delta r = 100\,$pc, and that the local density is that of an isothermal sphere with circular velocity $v$ at radius $R=10\,$kpc, $\rho = v^2/(4\pi G R^2)$ then we find that the overall dispersion in velocities becomes $\Delta v = N^{1/2}\delta v \approx 1\,$km/s.

This estimated magnitude is interesting because some tidal streams in our Galaxy are observed to have velocity dispersions of the same order, suggesting that these streams might be able to exclude (or detect) the perturbations predicted to arise in FDM halos.  
Tidal streams are arc-like structures consisting of stars that have been tidally stripped from star clusters or satellite galaxies moving through the tidal gravitational field of a larger, host galaxy.  If the original star cluster is dynamically cold, with a low escape velocity, then the stripped stars will move on very similar orbits as the original cluster, but with a slightly different orbital frequency that causes them to drift away from the original cluster in nearly the direction of the cluster's motion \cite{Johnston98a,Eyre11a,Bovy14a}.  Over time, the escaped stars trace out a nearly 1-dimensional structure called a tidal stream, whose width is related to the velocity dispersion of the stars.  Many observed streams have lengths vastly exceeding any possible FDM de Broglie wavelength in our Galaxy \cite{Pal5,GD1,Grillmair09a,ObservationalStreamsReview}, meaning that FDM perturbations do not act coherently across the entire stream, and therefore can generate a velocity dispersion in stream stars.
Some streams in the Milky Way, in particular the GD-1 stream \cite{Koposov10a,Bonaca2019,Bonaca20a}, are observed to have extremely small velocity dispersions of order $1\,\mathrm{km\,s}^{-1}$, similar in size to the ballpark estimate above for the expected magnitude of FDM-induced velocity dispersion.  
This motivates a more careful calculation of FDM perturbations on realistic tidal streams, which we discuss next.

\section{Method}

The most obvious way to calculate FDM effects on tidal streams would be to simulate a Milky Way-sized FDM halo, by numerical solution of the combined Schr\"odinger-Poisson equations using one or more of the algorithms published in the literature \citep[e.g.][]{Schive:2014dra,Li:2018kyk}.  While simulations are now straightforward for galaxies much smaller than the Milky Way, simulation of halos with mass $M\sim 10^{12}M_\odot$ can be onerously expensive. This is because of the requirement that the spatial grid used in the simulation must resolve the de Broglie wavelength, while the box must be larger than the halo being simulated.  The halo virial radius scales with halo mass as $R_{\rm vir}\propto M^{1/3}$, while the de Broglie wavelength scales as $\lambda \propto 1/v_{\rm vir}\propto M^{-1/3}$, meaning that the required number of spatial pixels scales with halo mass as $N\propto (R_{\rm vir}/\lambda)^3 \propto M^2$.  In addition, the Courant condition implies that the number of timesteps scales with halo mass like $v_{\rm vir}/\lambda\propto M^{2/3}$. 
Therefore, the computational expense of simulating halos of mass $M$ grows more quickly than $M^{8/3}$, meaning that even modest increases in halo mass require huge increases in computational resources.

\begin{figure*}
    \centering
    \includegraphics[width=0.4\textwidth,trim=1.5in 3in 1.5in 3in]{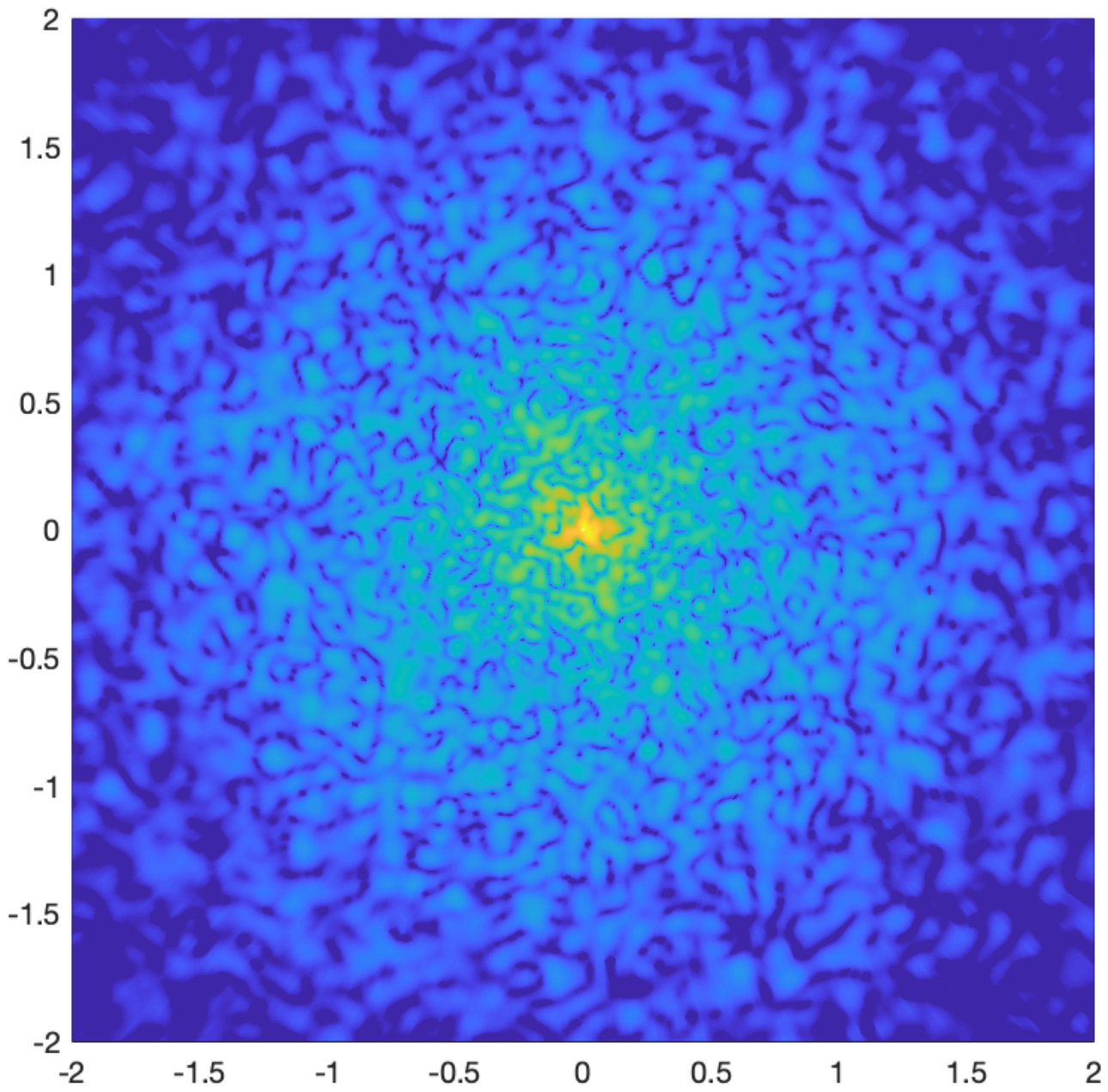}
    \includegraphics[width=0.58\textwidth]{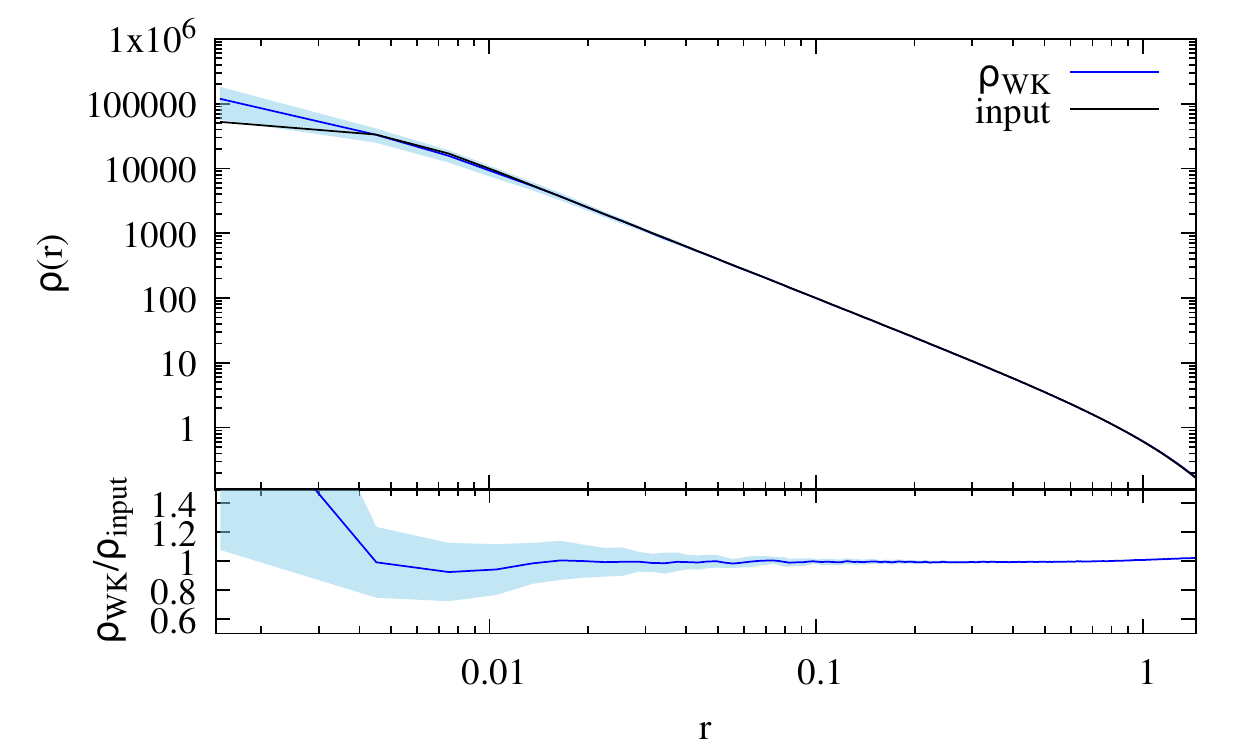}
    \caption{Examples of Widrow-Kaiser wavefunctions generated using Eqns.\ \eqref{decomp} and \eqref{WK}.  (Left) A slice through a realization with de Broglie wavelength $\lambda=h/mv_c = 0.1$ in code units.  The color scale is logarithmic in the density.  (Right)
    The black curve shows the input profile, and the blue curve shows the spherically averaged density profile $\rho=m|\psi|^2$, in units where the de Broglie wavelength is $\lambda=0.02$.  The shaded area shows the rms scatter in the density profiles over time.  The bottom panel shows the ratio.}
    \label{fig:density}
\end{figure*}

For this reason, we adopt an alternative approach which should provide an accurate calculation of the gravitational fluctuations of FDM halos.  We start by noting that the inner regions of MW-like halos where tidal streams are found (e.g., $R\lesssim 30\,$kpc) have dynamical times far shorter than the Hubble time, and so to a first approximation can be assumed to be in local equilibrium.  We therefore assume that we can describe the gravitational potential $\Phi(\bm{x},t)$ as a nearly static background potential $\Phi_0(\bm{x})$, along with small FDM perturbations that fluctuate over time.
This assumption is justified by previous 
simulations that have shown that after formation, isolated FDM halos maintain nearly static radial potential profiles, with significant temporal variations only close to the soliton core \citep{Li:2020ryg}.
We therefore compute FDM perturbations to linear order in perturbation theory, taking the smooth static potential as the zeroth-order solution.  If the zeroth-order potential is static, then the Hamiltonian governing the dynamics of the wavefunction has no explicit time dependence, meaning that the eigenmodes of the Hamiltonian evolve trivially over time as $e^{-i\omega t}$, where $\omega=E/\hbar$ and $E$ is the energy eigenvalue for each eigenmode.  We therefore wish to write the FDM wavefunction as a sum over eigenmodes of the zeroth-order Hamiltonian, 
\begin{equation}
\psi(\bm{x},t) = \sum_{i} a_{i} e^{-i \omega_i t} F_i(\bm{x}),
\label{decomp}
\end{equation}
where $i$ labels each eigenmode $F_i(\bm{x})$.  Given a solution for the coefficients $a_i$, evolving the wavefunction forward in time becomes a trivial calculation, given the eigenvalues $\omega_i$ and eigenfunctions $F_i(\bm{x})$ of the Hamiltonian.

In this paper, we will focus on spherical potentials, $\Phi_0(\bm{x})=\Phi_0(r)$, which makes the solution of the eigensystem of the Hamiltonian especially easy.  Just like in the hydrogen atom, we can decompose the eigenmodes into a product of radial functions $f_{nl}(r)$ and spherical harmonics $Y_{lm}(\theta,\phi)$; the only difference is that the potential is given by $\Phi_0(r)$ instead of the Coulomb potential. Solving for the eigenfunctions $f_{nl}$ and the eigenvalues $\omega_{nl}$ then is a simple exercise in linear algebra  \cite{Lin:2018whl}.  Our problem then reduces to picking the coefficients $a_{nlm}$ in Eqn.\ \eqref{decomp} that will not only give the desired density profile $\rho(\bm{x})=m |\psi(\bm{x})|^2$ on average, but that will also produce an equilibrium solution to Schr\"odinger-Poisson equations.  \citet{Widrow:1993qq} provide an elegant solution to this latter problem.  They propose the following ansatz for the wavefunction at an instant in time:
\begin{equation}
\psi(\bm{x}) = \sum_{\bm{p}} \sqrt{f(\bm{x},\bm{p}) d\bm{p}} N_{\bm{p}} e^{i \bm{x}\cdot\bm{p}},
\label{psi_WK}
\end{equation}
where the sum is over discrete momentum states spaced by $d\bm{p}$, $N_{\bm{p}}$ are independent random complex numbers with unit variance, and $f(\bm{x},\bm{p})$ is the classical distribution function that self-consistently solves the coupled Vlasov-Poisson equations for the desired potential $\Phi_0(r)$.  
We can immediately see the reasoning behind this ansatz by  computing the expectation value of the density $\rho=m|\psi|^2$. In the expectation value, the cross-terms multiplying different momentum states $\bm{p}$ and $\bm{q}$ vanish since $\langle N_{\bm{p}}^* N_{\bm{q}}\rangle=0$ for $\bm{p}\neq\bm{q}$, leaving behind only $\sum_{\bm{p}} \langle |N_{\bm{p}}|^2\rangle f(\bm{x},\bm{p}) d\bm{p} = \sum_{\bm{p}} f(\bm{x},\bm{p}) d\bm{p}$.  In the continuum limit, the sum becomes an integral over momentum of the distribution function $f(\bm{x},\bm{p})$ which by definition gives the density $\rho(\bm{x})$.  This ansatz therefore gives a simple solution to the problem of choosing the coefficients $a_{nlm}$ in Eqn.\ \eqref{decomp}: we construct the initial wavefunction according to Eqn.\ \eqref{psi_WK}, then project onto the eigenfunctions of the Hamiltonian $F_{nlm}(\bm{x})$ to determine the coefficients $a_{nlm}$, and then use Eqn.\ \eqref{decomp} to evolve the wavefunction over time.  To initialize the wavefunction, we need only determine the classical distribution function $f(\bm{x},\bm{p})$ to construct equilibrium solutions for the wavefunction, and for spherical potentials $\Phi_0(r)$ and isotropic velocity distributions, the famous Eddington formula provides a closed-form expression for the self-consistent ergodic distribution function \cite{BinneyTremaine}. 

In practice, we have used an even simpler approximation to the above construction.  Because we focus on spherical potentials and assume isotropy, the distribution function depends only on energy, $f(\bm{x},\bm{p})=f(E)$, and recall that the $F_{nlm}$ are eigenfunctions of the Hamiltonian whose eigenvalues $E_{nl}$ correspond to the energy of each state.  Therefore, 
instead of projecting Eqn.\ \eqref{psi_WK} onto the eigenfunctions, we directly write the coefficients $a_{nlm}$ in terms of the distribution function, using
\begin{equation}
a_{nlm}\propto\sqrt{f(E_{nl})} N_{nlm}.
\label{WK}
\end{equation}
Just as in Eqn.\ \eqref{psi_WK}, the $N_{nlm}$ are independent random complex numbers with unit variance.  
Figure \ref{fig:density}  shows an example of the density profile generated using Eqns.\ \eqref{decomp} and \eqref{WK}.  Evidently, this method does produce wavefunctions with density profiles closely matching the desired $\rho(r)$ profiles.  We find significant disagreement between the input profile and the reconstructed profile at $r\ll \lambda$, which is to be expected since the number of independent states at $r \lesssim \lambda$ becomes so small that the continuum limit of the distribution function becomes invalid.  We also find mild disagreement in cases where the input density profile declines very steeply at large radii.  We speculate that this occurs because the eigenmodes with support at these large radii are only marginally bound, and therefore cannot produce steep derivatives.

We have also verified that this method gives equilibrium solutions to the Schr\"odinger-Poisson equations, by evolving a small ($M\sim 10^{10} M_\odot$) halo generated using the Widrow-Kaiser ansatz Eqn.\ \eqref{psi_WK} with the simulation code of \cite{Li:2018kyk}.  The halo profile remains essentially unchanged over time, except for the rapid formation of a central soliton near $r=0$.  We could, in principle, amend this procedure by adding a soliton to our input profile, according to the emprical soliton mass -- halo mass relation found in cosmological simulations \cite{Schive:2014hza}.  However, since the calculations below will focus on distances far from the central soliton, which  should have a size $R_s \sim 0.1\,$kpc for galaxies like the Milky Way,
we ignore this one aspect in which our method disagrees with full nonlinear Schr\"odinger-Poisson simulations. 

For the calculations below, we will use a zeroth-order profile that is close to isothermal, $\Phi_0\sim v_c^2 \log r$, which has a density profile $\rho\propto r^{-2}$, consistent with the rotation curve for our Galaxy.  To make the total mass finite, we truncate the profile at radius $s$ by multiplying by $\exp(-r^2/2s^2)$.  Since $|\psi^2|$ cannot support a power-law divergence at $r\ll \lambda$, we also add a core radius, by subtracting a Gaussian with scale radius $c$.  So our zeroth-order density profile is 
\begin{equation}
\rho_0(r) = \frac{v_c^2}{4\pi G\,r^2} \left[
\exp\left(-\frac{r^2}{2s^2}\right)-\exp\left(-\frac{r^2}{2c^2}\right)
\right]
\label{eqn:rho0}
\end{equation}
We will generally focus on observables at locations $c\ll r\ll s$, in which case the precise choices for the parameter values of $s$ and $c$ do not appear to affect any of the results.

\begin{figure}
    \centering
    \includegraphics[width=0.48\textwidth]{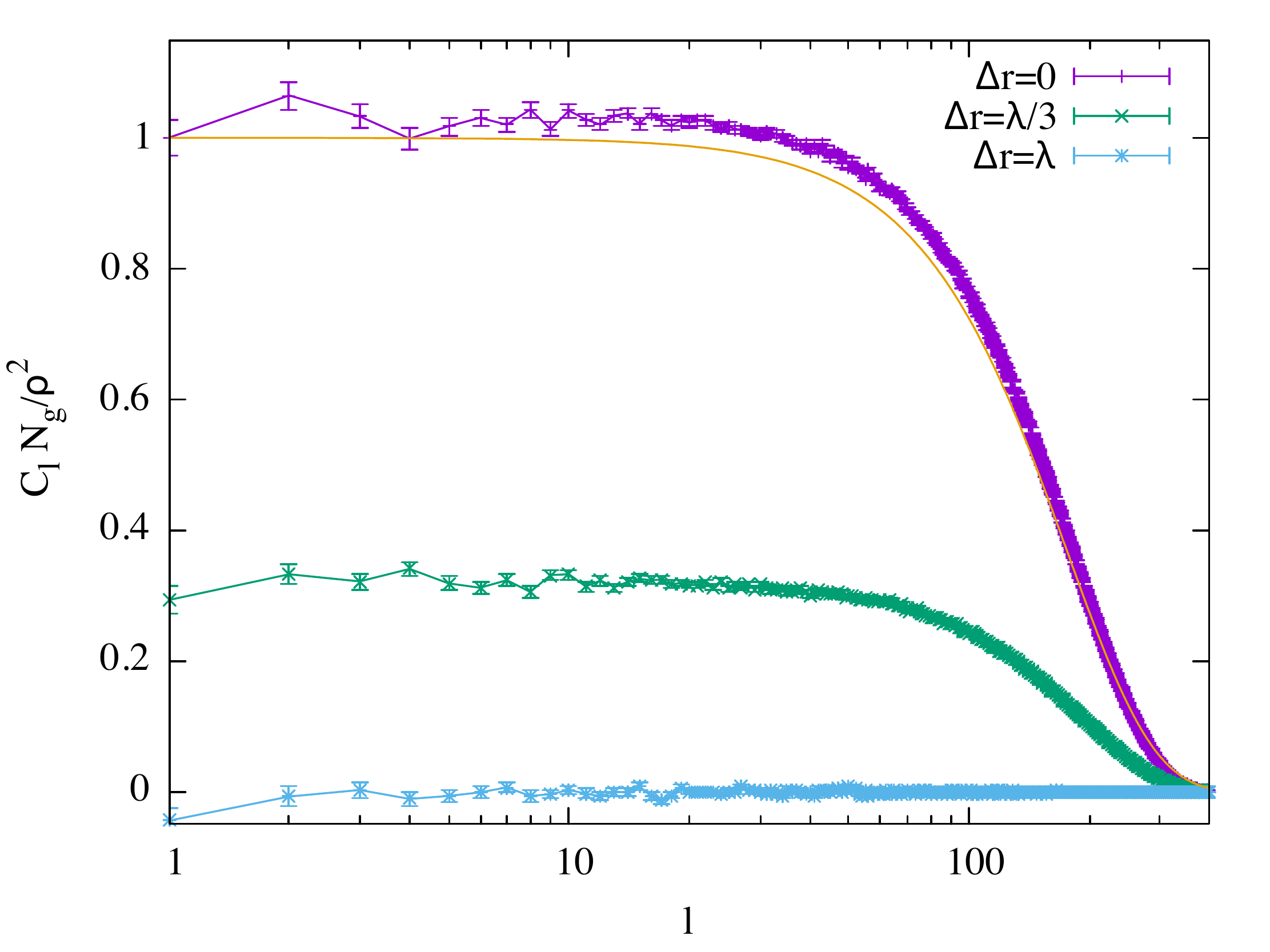}
    \caption{Angular power spectrum of FDM density fluctuations for the same halo shown in Fig.\ \ref{fig:density}, evaluated at $r=25\lambda$, for shells separated by various $\Delta r$.  The auto-spectrum ($\Delta r=0$) is well described as white noise on large angular scales, with an exponential damping on small scales. The thin yellow line shows $C_l=({\bar\rho}^2(r)/N_g) \exp(-a [l \lambda/2\pi r]^2)$, which is a reasonable approximation for $a=0.8$.  
    Shells at different radii become uncorrelated at separations $\Delta r > \lambda/2\pi$, as shown by the green and blue curves.}
    \label{fig:cl_rho}
\end{figure}

Fig.\ \ref{fig:density} shows an example of this profile, and a realization of the corresponding Widrow-Kaiser wavefunction generated using Eqns.\ \eqref{decomp} and \eqref{WK}.  As the figure shows, the FDM density exhibits density fluctuations from wave interference.  We can quantify the statistics of these fluctuations using their two-point correlations.  From the spherical harmonic coefficients $\rho_{lm}(r)=\int \rho(r,\theta,\phi) Y_{lm}^*(\theta,\phi) d\cos\theta d\phi$, let us define the angular cross-power spectrum between the density at radial shells $r$ and $r^\prime$ as
\begin{equation}
    \langle \rho_{lm}^*(r) \rho_{l^\prime m^\prime}(r^\prime)
     = C_l(r,r^\prime) \delta_{l l^\prime} \delta_{m m^\prime}.
\end{equation}
Figure \ref{fig:cl_rho} plots $C_l$ for the same halo shown in Fig.\ \ref{fig:density}.  The auto-spectrum of FDM density fluctuations appears consistent with white noise on large angular scales that becomes exponentially damped on small scales, well described by a simple form, 
\begin{equation}
   C_l(r) \approx \frac{{\bar\rho}^2(r)}{N_g} 
   e^{-a \left(l\lambda/2\pi r\right)^2},
   \label{eqn:whitenoise}
\end{equation}
where ${\bar\rho}(r)$ is the mean density at radius $r$,  $N_g=4r^2/\lambda^2$ is the number of independent `granules' of cross-sectional area $\pi \lambda^2$ in the surface area $4\pi r^2$ of a sphere of radius $r$, and we find that $a\approx0.8$ better describes the location of the exponential damping than $a=1$.  Density fluctuations at different radii become uncorrelated with each other once their separation $\Delta r > \lambda/2\pi$.  

These FDM fluctuations can perturb the motions of stars in tidal streams.  To compute the effect on tidal streams, we must solve the Poisson equation for the perturbed gravitational potential $\Phi(\bm{x},t)$ given the perturbed density $\rho$. Since we have written the eigenfunctions in terms of spherical harmonics, this is easily accomplished using fast spherical harmonic transforms (SHT).  At each timestep, we evaluate Eqn.\ \eqref{decomp}, square $\psi$ in configuration space to obtain $\rho(\bm{x},t)$ and then compute spherical harmonic coefficients $\rho_{lm}(r,t)$ by SHT using the \texttt{SHTns} \footnote{\url{https://nschaeff.bitbucket.io/shtns/index.html}} library \cite{shtns}.  Given $\rho_{lm}(r)$, we can solve for the potential $\Phi_{lm}(r)$ in the usual way, using variation of parameters:
\begin{eqnarray}
\Phi_{lm}(r) = -\frac{4\pi G}{2l+1} &&\left[
r^l \int_r^\infty R^{1-l} \rho_{lm}(R) dR  \right.  \\
&&+  \left.
r^{-(l+1)}\int_0^r R^{2+l}\rho_{lm}(R) dR
\right].   \nonumber
\label{eqn:poisson}
\end{eqnarray}
Given $\Phi_{lm}(r,t)$, we then compute the gravitational acceleration $-\nabla\Phi(\bm{x},t)$ in configuration space  by SHT.

To summarize, in this section we have presented a simple and inexpensive method to evolve realistic wavefunctions for bound, virialized systems.  As mentioned above, this method should be accurate to first order in the potential perturbations.  Since the potential perturbations are indeed small for the regime of interest (streams at distances $R \sim 10-20\,$kpc in halos with $\lambda/2\pi \sim 0.1\,$kpc), linear order should be adequate for our intended calculation.  We stress the computational simplicity of this method: rather than requiring giant supercomputers, the simulations discussed below were all performed on a single compute node.

\section{Stream perturbations}
\label{sec:perts}

\begin{figure}
    \centering
    \includegraphics[width=0.48\textwidth]{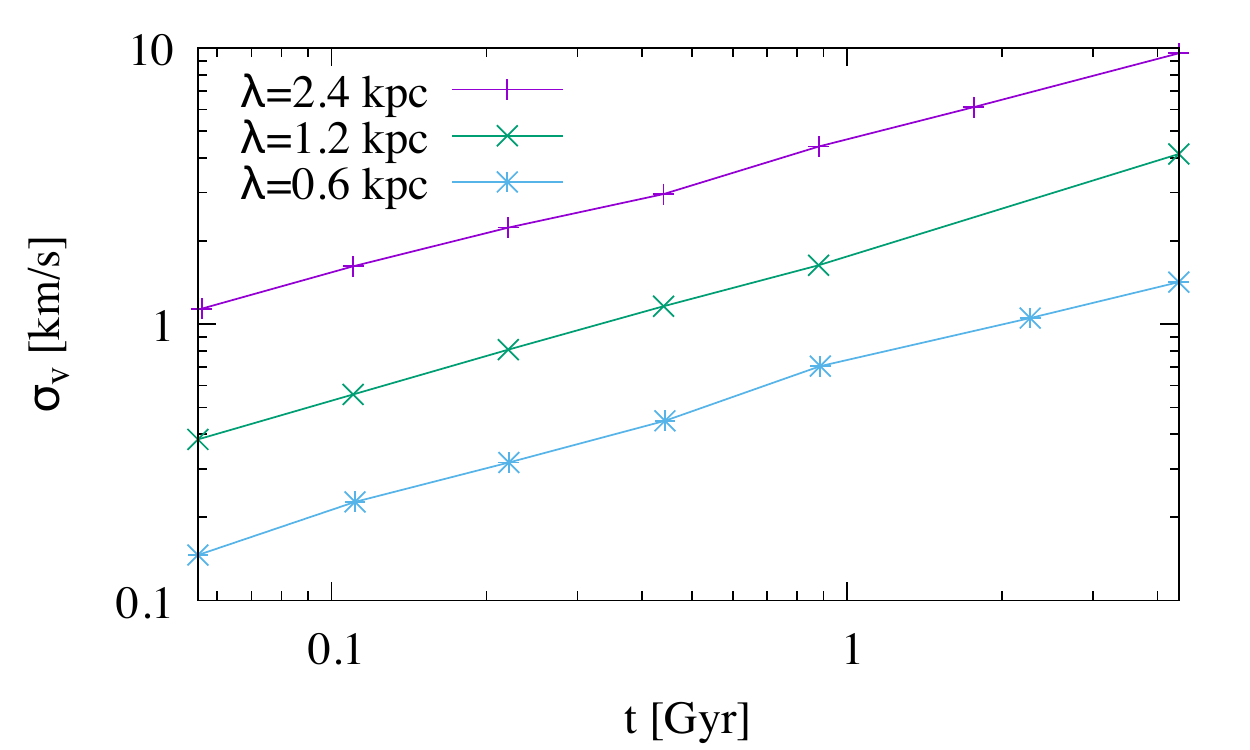}
    \caption{Growth of the velocity dispersion over time.  The various curves show the velocity dispersion for different values of the de Broglie wavelength $\lambda$.  The dispersion agrees well with the order-of-magnitude estimate described in \S\ref{sec:intro}, approximately scaling as $\sigma_v^2\propto \lambda^3 t$.}
    \label{fig:variance}
\end{figure}

In this section, we show example results from simulations of stellar streams moving through FDM halos evolved using the method described in the previous section.  In all simulations discussed below, the star particles are treated as test particles that respond to gravity but do not generate their own gravitational forces. The FDM halo profile is chosen to be an isothermal sphere with circular velocity $v_c=200\,$km/s, which gives a de Broglie wavelength $\lambda = h/(m\,v_c) = 0.6\,$kpc for the canonical FDM mass $m=10^{-22}\,$eV.  For simplicity, we study streams on initially circular orbits at a radius $r_0=15\,$kpc.  Because the star particles never reach radii significantly larger than $r_0$, we do not compute the potential out to the virial radius, but instead truncate the computational volume at a radius $r_{\rm max}=90\,$kpc.  As long as $r_{\rm max}$ significantly exceeds $r_0$, we find that the simulation results are insensitive to this choice.  We solve for eigenfunctions on a radial grid spaced linearly in radius over the range $0<r<r_{\rm max}$, typically using $n_r=1000$ radial grid cells, so that the grid spacing $r_{\rm max}/n_r \lesssim \lambda/2\pi$.  We include spherical harmonics up to $l_{\rm max}=400$.  From comparison with a few test cases with $n_r=2000$ and $l_{\rm max}=600$, it appears that our fiducial choices for $n_r$ and $l_{\rm max}$ are adequate to achieve convergence in the perturbation spectra discussed below.

As a first example, we show in Figure \ref{fig:variance} the velocity dispersion induced by the FDM perturbations.  To calculate this, we initialize test particles along a circular ring at radius $r_0=15\,$kpc, moving on initially circular orbits with identical angular momenta.  The particles then evolve in the fluctuating FDM potential, and we plot the  velocity dispersion, $\sigma_v=\sigma_L/r_0$, where $\bm{L}=\bm{r}\times\bm{v}$ is the angular momentum vector for each particle, and $\sigma_L^2 = (\langle L^2\rangle-|\langle \bm{L}\rangle|^2)/3$.  As expected, the variance scales like $\sigma_v^2 \propto \lambda^3 t$, growing linearly in time as particles diffuse away from their initially circular orbits.  Note also that the amplitude of $\sigma_v$ is similar to the order-of-magnitude estimate given in \S\ref{sec:intro}.

\subsection{Power spectra}
\label{sec:powerspectra}

\begin{figure}
    \centering
    \includegraphics[width=0.48\textwidth]{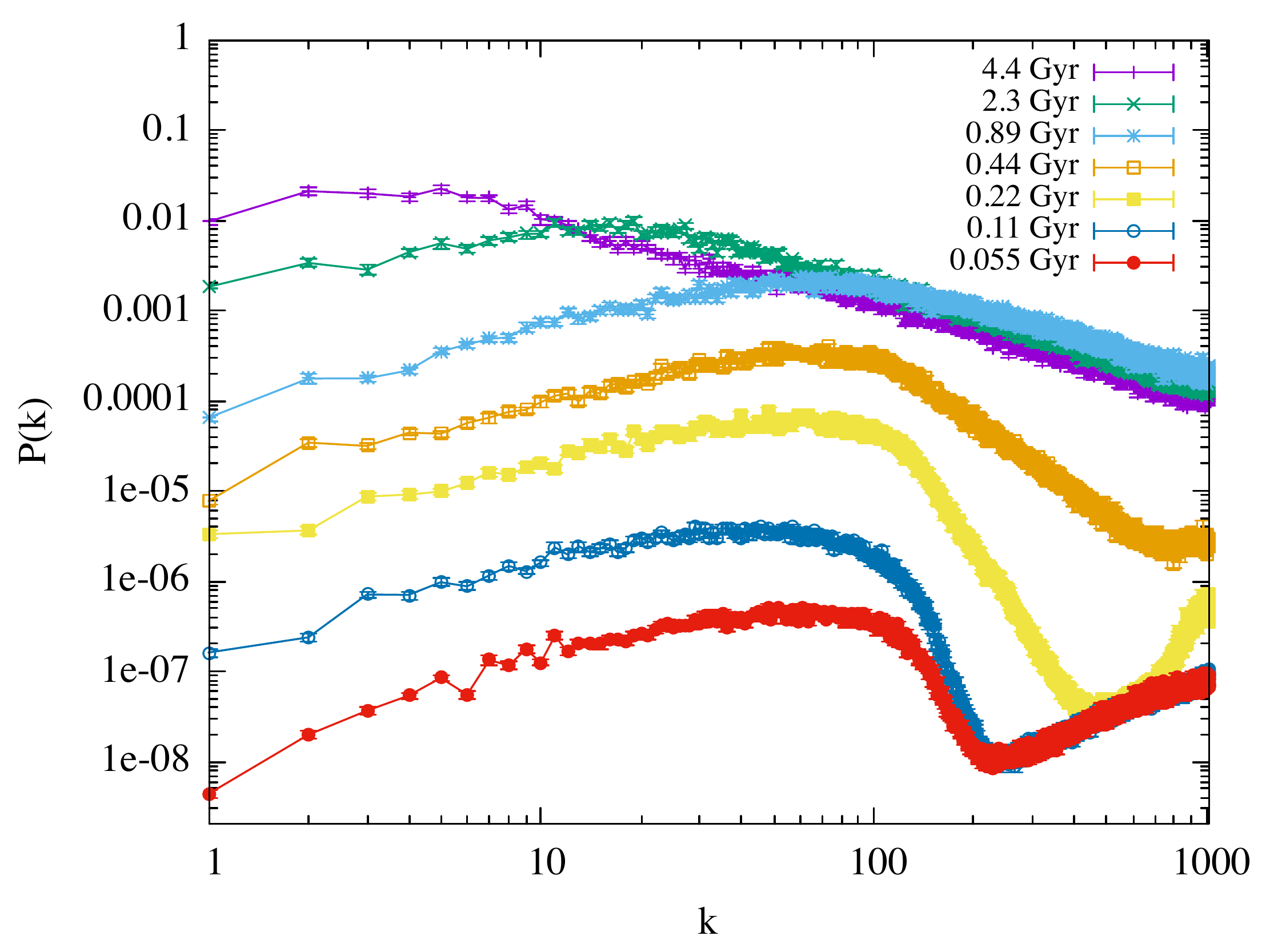}
    \caption{Density power spectra for a stream of $N_p=10^5$ stars at $r_0=15\,$kpc in an isothermal halo with $v_c=200\,$km/s and $m=10^{-22}\,$eV, giving 
    $\lambda=0.6$ kpc.  The cutoff in $P(k)$ near $k\sim 100$ is due to the cutoff in the FDM angular power spectrum at $k=2\pi r_0/\lambda$; the excess power at $k\gtrsim 200$ in the bottom curves is spurious and arises from  discreteness noise.  Note that at times $t\gtrsim 1\,$Gyr the power spectrum at high $k$ has become a power law $P \propto k^{-1}$, and ceases to grow in amplitude over time.}
    \label{fig:ring_pk}
\end{figure}

\begin{figure*}
    \centering
    \includegraphics[trim=1in 0 1in 0,width=0.36\textwidth]{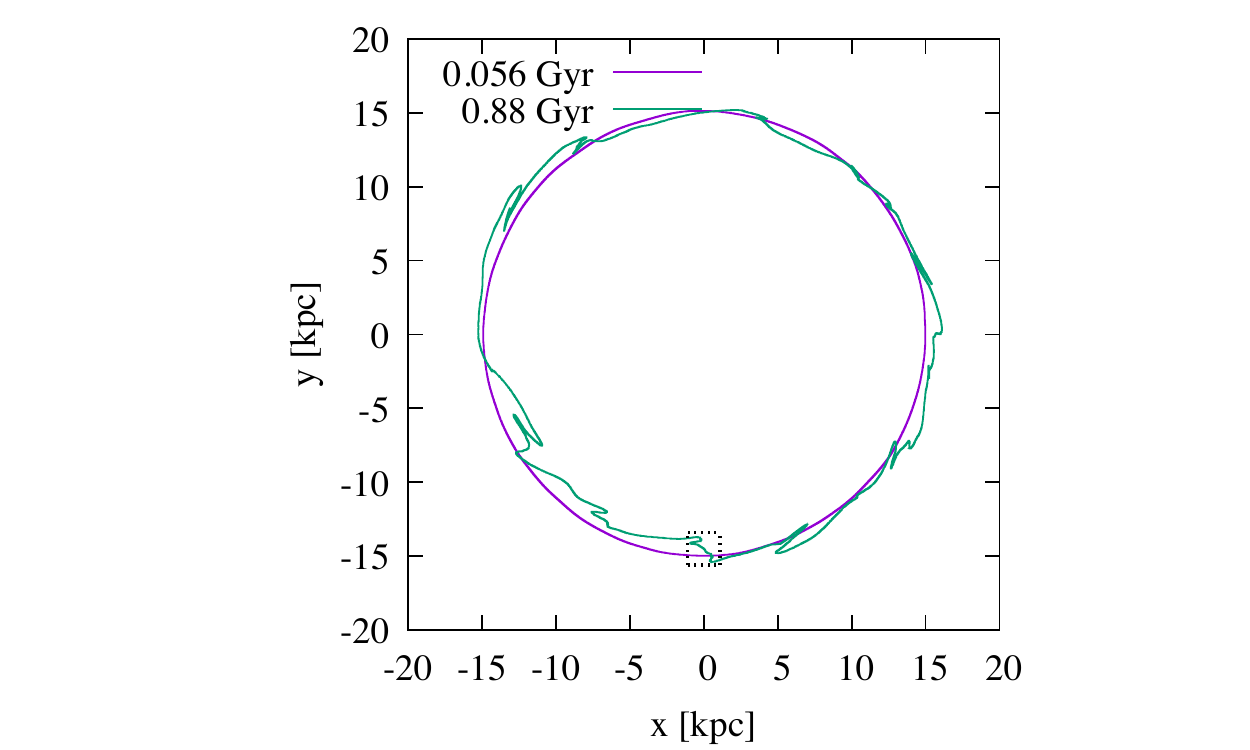}
    \hfil
    \includegraphics[width=0.6\textwidth]{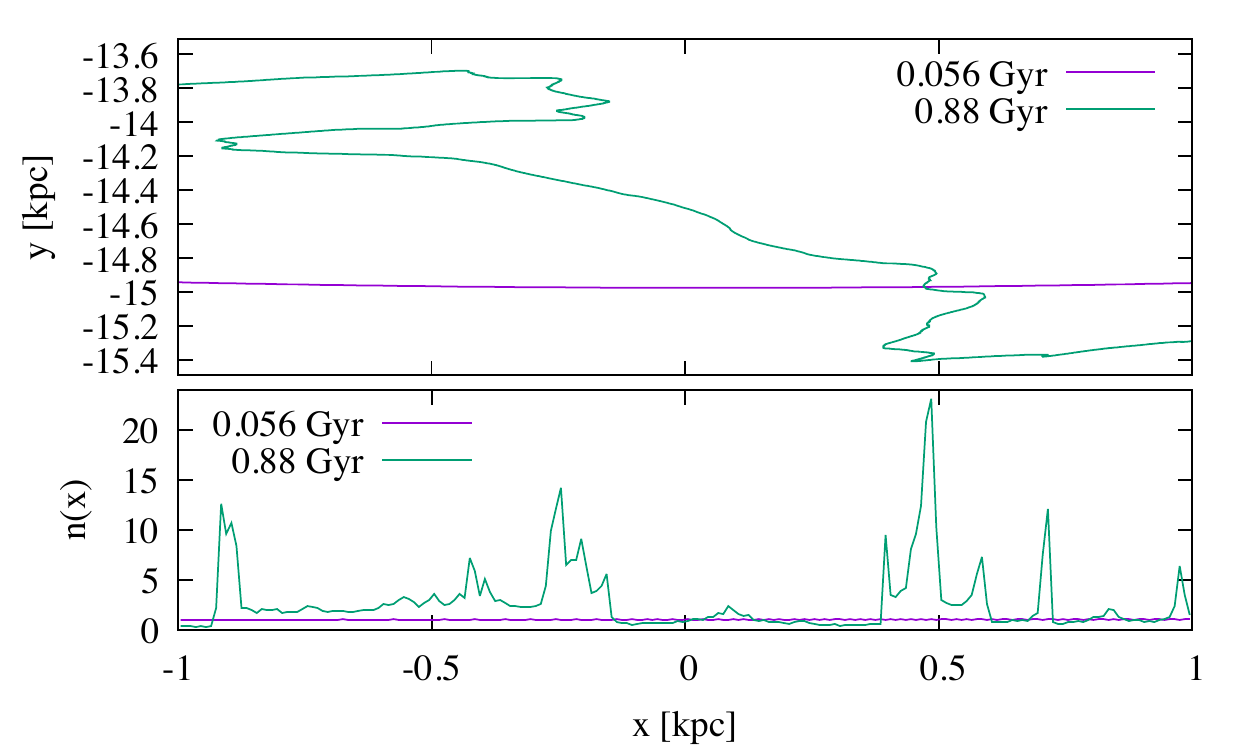}
    \caption{Examples of fold caustics.  (Left) The two curves show streams evolved in a simulation with $\lambda = 2.4\,$kpc.  Over time, the stream folds on itself, producing density variations on scales much smaller than $\lambda$. (Right) The upper panel zooms in on the region enclosed by the dotted black square in the left panel.  The bottom panel shows the binned 1D number density of particles; note the large spikes in density at fold locations in the upper panel, reflecting the universal $n\propto x^{-1/2}$ divergence at fold caustics in 1D.}
    \label{fig:fold}
\end{figure*}

The total variance does not fully capture all aspects of the FDM perturbations.  In particular, because the density field is coherent on scales below the de Broglie wavelength, we might expect that the stream perturbations will similarly show spatial coherence.  
To quantify this, we measure the 1D density power spectrum along the stream, which previous work has shown to be sensitive to the mass scale and length scale of potential fluctuations in CDM halos \cite{Bovy2017}.
We compute this 1D power spectrum by counting the number of test particles found in bins of azimuthal angle $\phi$ in the mean orbital plane, $n(\phi)$, dividing by the mean $\langle n\rangle$, and then Fourier transforming.  We normalize the power spectrum so that $P=1$ at $k=0$, and a uniformly random Poisson distribution of $N_p$ particles has an average power spectrum $\langle P\rangle=N_p^{-1}$ at $k>0$.

We first show power spectra for simulations initialized with test particles placed uniformly along a circular ring, moving on exactly circular orbits.  In the absence of FDM perturbations, the particles remain on exactly circular orbits uniformly spaced in azimuthal angle $\phi$, so that the 1D density power spectrum vanishes on scales below the Nyquist frequency.  In FDM halos, the test particles are perturbed and their density power spectrum becomes nonzero.  Fig.\ \ref{fig:ring_pk} shows the time evolution of the 1-D power spectrum for a ring initially at $r_0=15\,$kpc, evolving in a halo with $\lambda=0.6\,$kpc.  
At early times, $t\lesssim 0.1\,$Gyr, the stream power spectra in Fig.\ \ref{fig:ring_pk} 
behave like $P(k) \propto t^3 k$ for $k \ll 2\pi r/\lambda$. This $t^3 k$ scaling is easy to understand, and is explained in Appendix \ref{sec:pre}.  
On smaller scales, $k>2\pi r/\lambda$, the power spectra initially exhibit a sharp cutoff.  This damping is to be expected: if the gravitational forces are smooth on small scales, then the displacement perturbations and density perturbations that they generate will also be smooth on small scales.  However, at subsequent times the cutoff disappears, and instead the power spectrum becomes a power-law $P(k)\propto k^{-1}$ at high $k$.  In addition, at late times $t\gtrsim 1\,$Gyr, the power spectrum stops growing at high $k$, and indeed slightly shrinks in amplitude after $\sim 2\,$Gyr.  This late-time saturation of the 1D power spectrum may seem surprising, since FDM fluctuations continue to scatter the motions of stars at all times.

The reason for this behavior is that the 1D power spectrum becomes dominated by the formation of fold caustics along the stream.  Fold caustics are the simplest type of catastrophe \citep{Arnold1984}, and occur in a variety of settings, including gravitational lensing \citep{Blandford1986}, dark matter halos \citep[e.g.][]{Fillmore1984,Diemer2014,Adhikari2014}, and tidal streams \cite{Carlberg2009,Erkal2015}.
Because the FDM perturbations act coherently, entire sections of the stream can overtake other sections, causing the stream to fold over on itself.  Figure \ref{fig:fold} shows an example.  The time when folds first appear scales with de Broglie wavelength as $t\propto \lambda^{-1/3}$, as explained in Appendix \ref{sec:pre}.  For our fiducial case of a circular stream at radius $r_0=15\,$kpc in an isothermal halo with $v_c=200\,$km/s and $m=10^{-22}\,$eV, this time is $t\sim 0.7\,$Gyr.

The formation of fold caustics significantly affects the 1D density along the stream.   If the stream initially has vanishing velocity dispersion (infinite phase space density), then the 1D density $n(x)$ diverges at the fold caustic as $\delta n \propto (r_c/x)^{1/2}$, where $x$ is the distance to the fold caustic, and $r_c$ is a coherence length of the fold.  The Fourier transform of this density profile behaves as $\delta n(k) \propto (r_c/k)^{1/2}$, and so each fold contributes to the power spectrum as $P\sim r_c/k$.  This explains why the power spectra shown in Fig.\ \ref{fig:ring_pk} transition to $P\propto k^{-1}$ after some time; this occurs when the stream perturbations become so large that the stream folds over on itself.

This also explains why the 1D density power spectra saturate in amplitude.  Because the stream has a finite length $L$, there is a maximum number of fold caustics of size $r_c$ that can occur in the stream, proportional to $L/r_c$.  When the stream becomes filled with caustics, any additional perturbations will destroy as many caustics as they create, leading to the total power spectrum saturating in amplitude on small scales.  On large scales, the power spectrum can continue to grow, as larger and larger caustics develop over time.  Eventually, the power spectrum asymptotes to a $k^{-1}$ profile, independent of the form of the gravitational fluctuations that produce the stream perturbations.  This behavior is quite easy to reproduce using simple toy examples, in which we randomly displace particles in a ring.  For small perturbations, the resulting power spectrum reflects the form of the displacement perturbations, but once the displacements become large enough that folds occur, the power spectrum exhibits the universal behavior discussed above.

\subsection{Effect of velocity dispersion}
\label{sec:dispersion}

\begin{figure}
    \centering
    \includegraphics[width=0.48\textwidth]{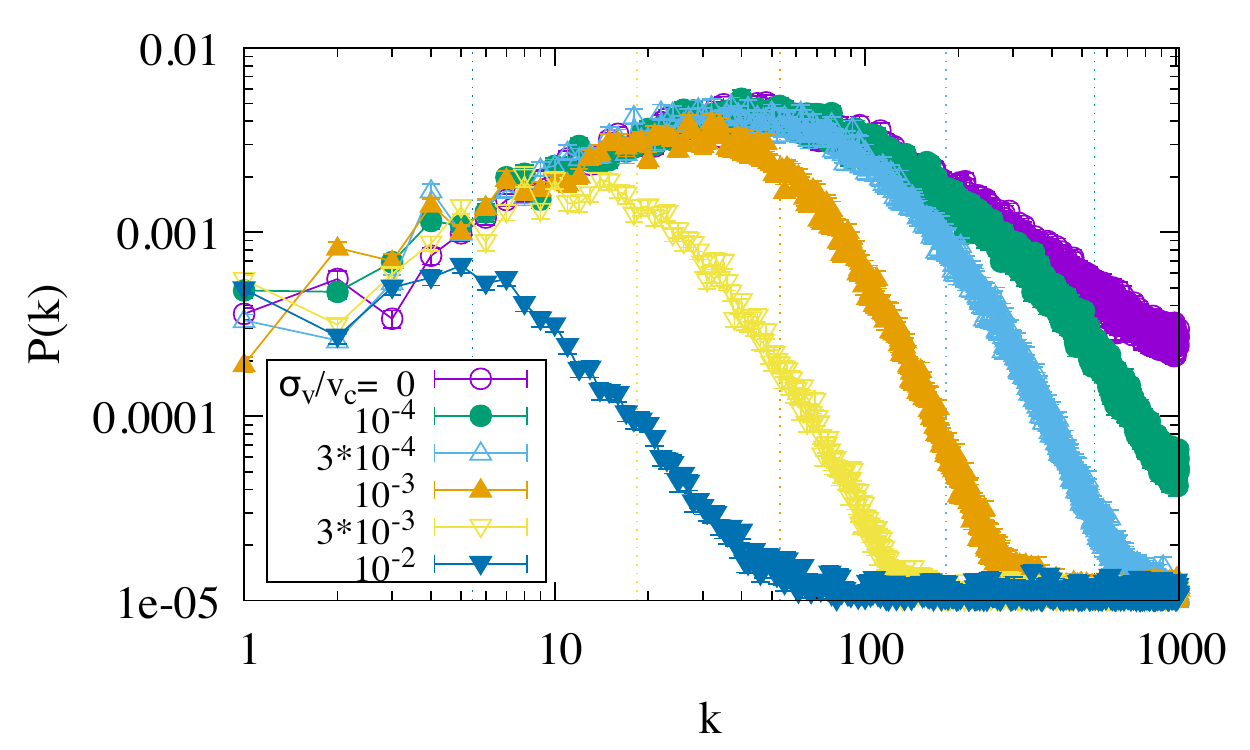}
    \caption{Effect of velocity dispersion on power spectra.  The curves show power spectra at $t\sim 1.3\,$ Gyr for $\lambda=0.6\,$kpc, and are labeled by the initial 1D velocity dispersion $\sigma_v$ of the test particles.  The vertical thin dotted lines show the location $k_d=r_0/(\sigma_v t)$, where we expect dispersion to damp the power spectrum by a factor $\approx e^{-1}$.}
    \label{fig:sigma}
\end{figure}

\begin{figure*}
    \centering
    \includegraphics[width=0.48\textwidth]{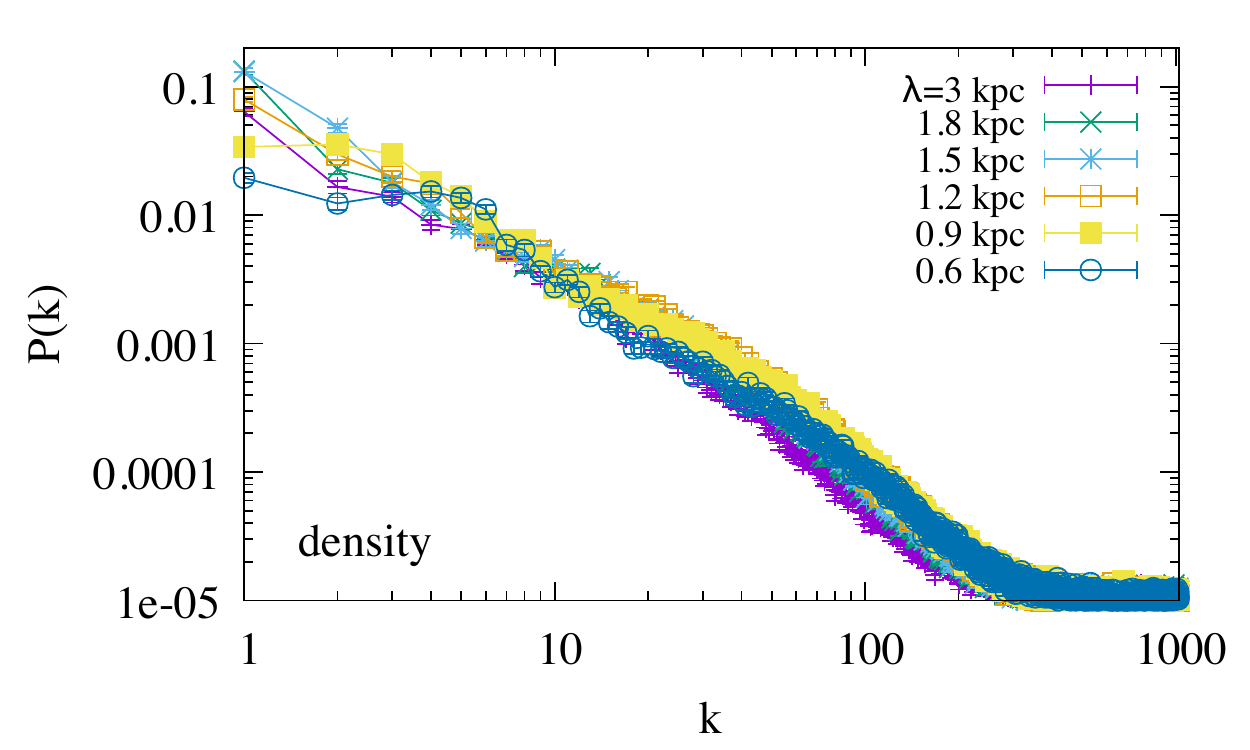}
    \hfil
    \includegraphics[width=0.48\textwidth]{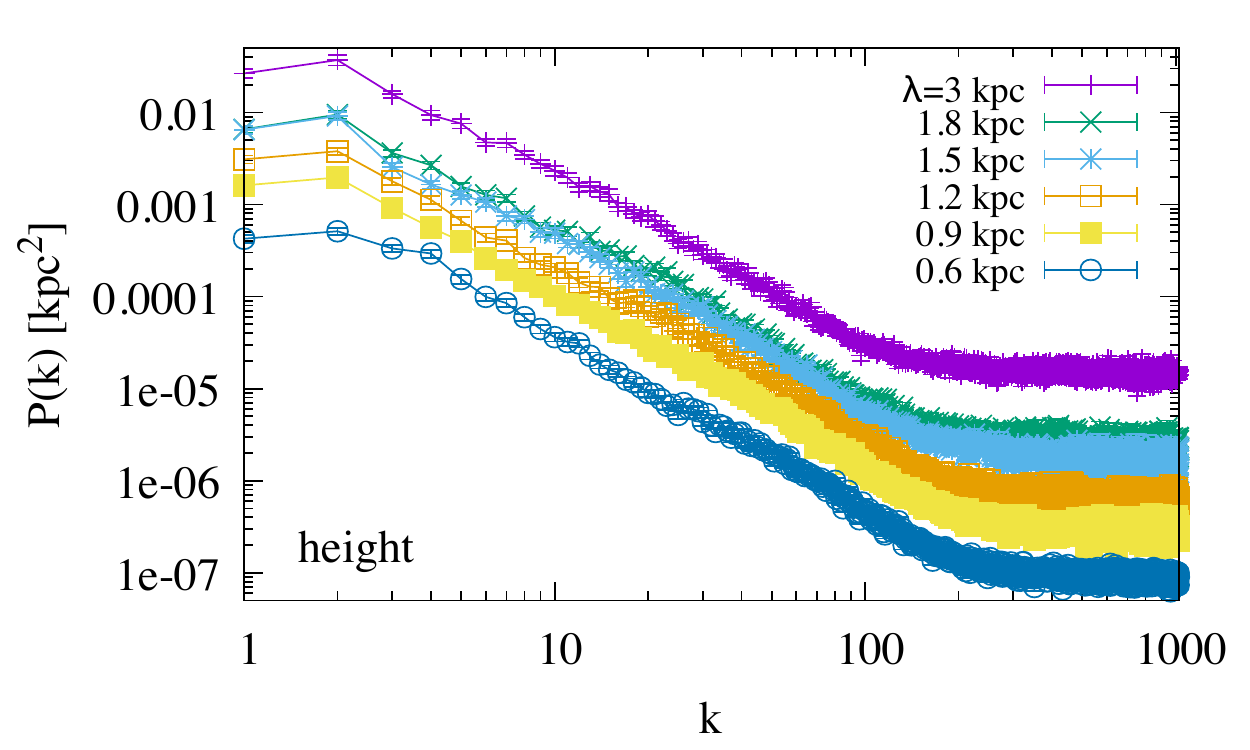}
    \caption{Power spectra as a function of de Broglie wavelength.  The left panel shows the overdensity power spectrum, and the right panel shows the power spectrum of height $z$ perpendicular to the mean orbital plane.  All simulations had an initial velocity dispersion $\sigma_v = 3\times 10^{-3} v_c = 0.6\,$km in each direction, and power spectra are shown after $t=4.4\,$Gyr of evolution.  Note that the amplitude of the  power spectrum of height fluctuations does not saturate in the same way that the 1D density $P(k)$ does.  Similar behavior is seen in the correlations of other quantities, like angular momenta or angular frequency.}
    \label{fig:zz}
\end{figure*}

This universal form of the power spectrum arises because of the $x^{-1/2}$ divergence near fold caustics.  As noted above, this divergence occurs only in the limit of infinite phase space density, i.e.\ vanishing velocity dispersion.  A finite dispersion in the stream regularizes the divergence, which changes the density profile at the caustic and hence also modifies the power spectrum.  Figure \ref{fig:sigma} illustrates this effect, showing power spectra for various values of the initial 1D velocity dispersion of the test particles.  Gaussian random velocities were added to all 3 components of the initial velocities, so that the 3D dispersion is $\sqrt{3}$ times the 1D dispersion, but the parallel component is the most important component.  Because we are considering small perturbations to circular orbits in an isothermal potential, a small change to the tangential velocity $\Delta v_\parallel$ changes the instantaneous angular velocity by $\Delta\omega=\Delta v_\parallel/r_0$, and the time-averaged angular velocity by $\Delta\Omega=-\Delta\omega$, so the dispersion in velocity $\sigma_v$ gives a dispersion in angular frequencies of $\sigma_\Omega=\sigma_v/r_0$.  This dispersion smears out any caustics by an angle $\sigma_\phi=\sigma_\Omega t$, which damps the power spectrum on scales smaller than $k_d = \sigma_\phi^{-1}$.  In addition to this damping, the velocity dispersion also adds shot noise from Poisson fluctuations in the star counts at a level $P=N_p^{-1}$ at $k>k_d$.

\subsection{Other correlations}
\label{sec:others}

From the above results, we see that at late times, the density power spectra of the streams in our FDM halos tend towards universal forms that depend mainly on the initial velocity dispersion, rather than the coherence length of the FDM fluctuations.  Although the de Broglie wavelength is initially imprinted in the shape of the density power spectrum, once fold caustics develop in the stream, both the amplitude and shape of the power spectrum become insensitive to $\lambda$.  This does not mean, however, that we cannot determine FDM parameters in this caustic regime.  The reason is that we can measure other properties of stars in addition to their number density.  Using precise astrometry from missions like Gaia \citep{GaiaMission}, along with radial velocity observations, it is possible to determine full 6-D phase space coordinates for stars in streams, from which other quantities may be derived, like their angular momenta or action-angle coordinates.  All of these quantities are perturbed by FDM fluctuations, and as we saw in Fig.\ \ref{fig:variance}, the variance of these perturbations does not saturate over time, unlike the 1D density power spectrum.  Fig.\ \ref{fig:zz} shows an example.  The left panel shows the 1D density power spectra for simulations with different values of $\lambda$, and we can see that eventually all the power spectra appear similar in both shape and amplitude.  The right panel shows the power spectrum of $z$, the coordinate perpendicular to the mean orbital plane of the stars.  We measure the power spectrum of $z$ by computing the average $z$ in bins of azimuthal angle $\phi$ in the mean orbital plane, and then Fourier transforming $z(\phi)$ along the $\phi$ direction.  In the caustic-dominated regime, the shape of the $z$ power spectrum is similar to the shape of the density power spectrum, but the amplitude depends sensitively on $\lambda$.  
If we can determine the age of the stream, then the amplitude of the $z$ power spectrum (or similarly, the correlations of other quantities like angular momenta or actions) can be used to determine the de Broglie wavelength, even when caustics dominate the power spectrum.  For the streams we have shown so far, which are initialized as circular rings, there is no obvious way to determine the stream age from the phase space data alone.  However, for realistic streams produced from stripping or disruption or star clusters, the length of the stream is related to the stream's age.  We therefore consider streams originating from compact clusters next.

\subsection{Streams from star clusters}

\subsubsection{Instantaneous escape}

\begin{figure}
    \centering
    \includegraphics[width=0.48\textwidth]{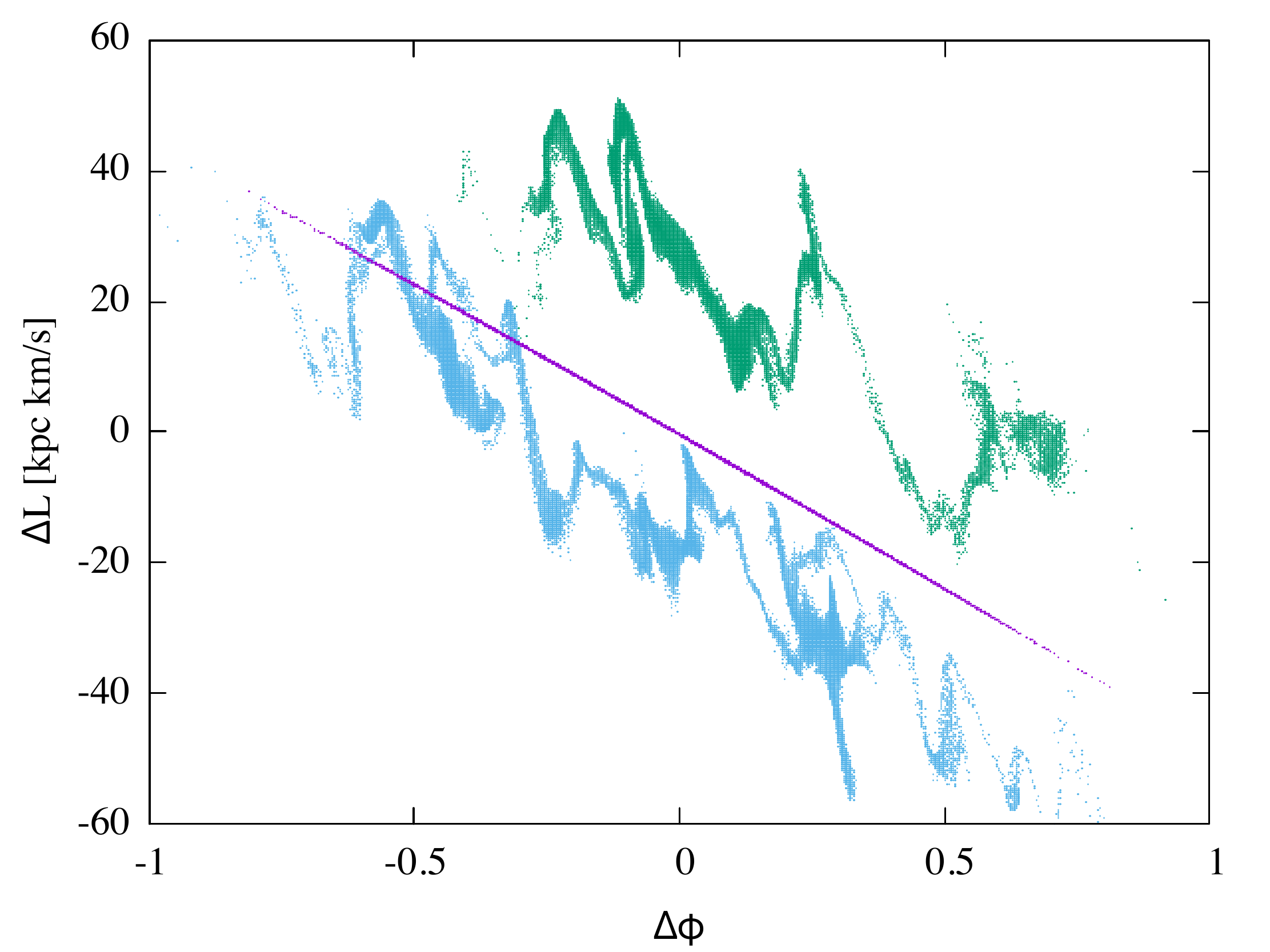}
    \caption{Examples of streams from compact clusters.  The x-axis is $\Delta\phi=\phi-\langle\phi\rangle$, and the y-axis is $\Delta L=L-L_0$, where $L_0$ is the mean specific angular momentum of the initial cluster.  The purple points correspond to a stream evolved for 4.4 Gyr in a smooth potential, while the green and blue points are from streams evolved in FDM potentials with $\lambda=0.6\,$kpc, also for 4.4 Gyr.  In all cases, the initial cluster of test particles had velocity dispersion $\sigma_v=0.6\,$km/s in each direction.}
    \label{fig:phiL}
\end{figure}

In the previous subsections, we showed simulations in which test particles were initialized on circular orbits spaced uniformly in a ring.  Next, we discuss simulations where test particles are initialized in a compact cluster.  The test particles initially are at the same location, at $r_0=15\,$kpc, moving on identical circular orbits. We then add Gaussian random velocities with dispersion $\sigma_v$ to all 3 components of the particles' velocities, and allow them to evolve.  The scatter in the particles' velocities gives a scatter in their angular frequencies $\Omega$, and so the stream lengthens over time, growing in length as $\propto t\, \Delta\Omega$.  
This occurs both in the FDM potential and also in a smooth potential.  In the smooth potential, the stream particles essentially sort themselves by $\Omega$ as time progresses, whereas in the FDM case, the fluctuations in the potential can act to jumble up the ordering of $\Omega$ as a function of $\phi$, as illustrated in Fig.\ \ref{fig:phiL}.

\begin{figure}
    \centering
    \includegraphics[width=0.48\textwidth]{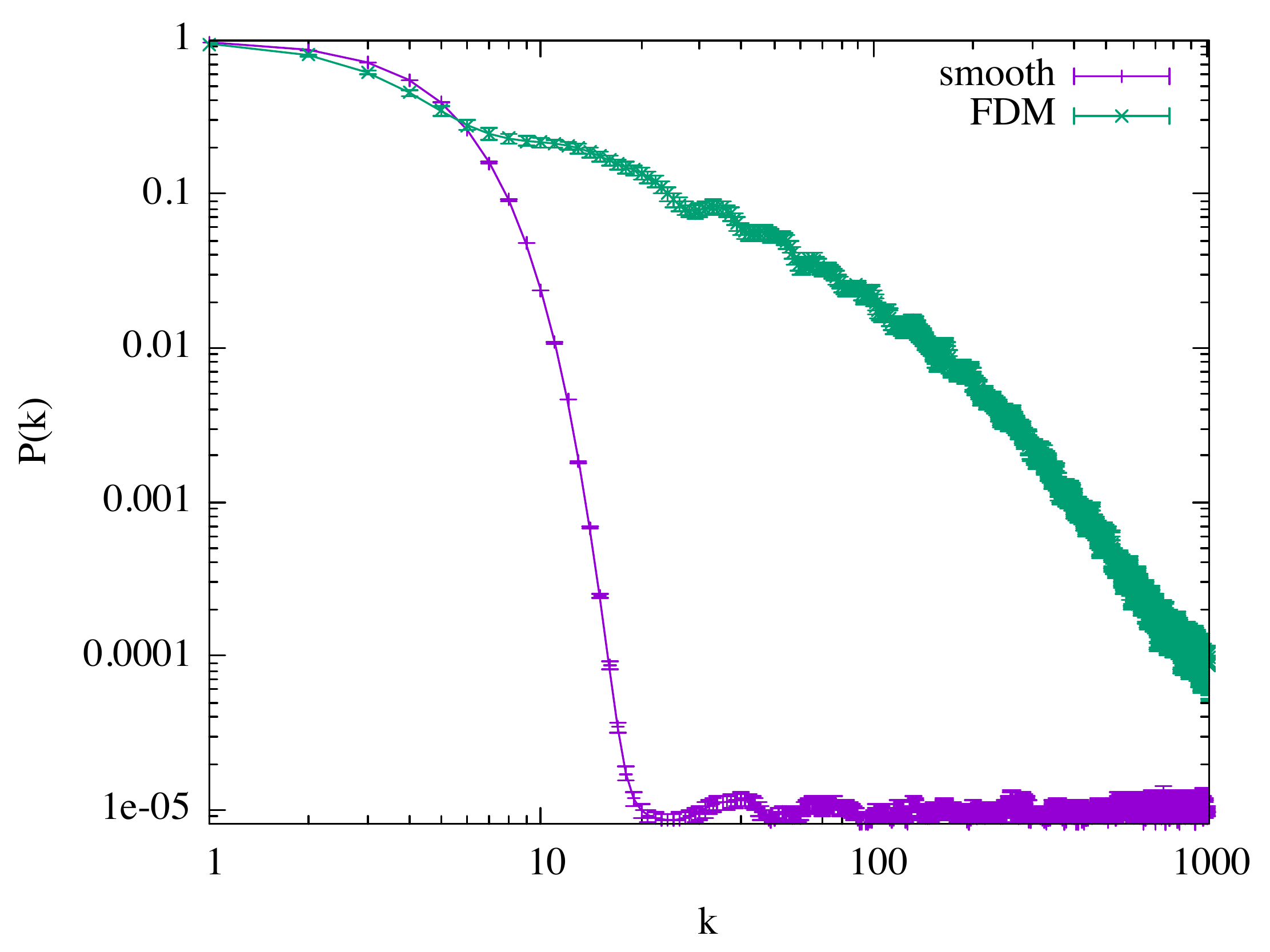}
    \caption{Density power spectra for streams originating from compact clusters of test particles.  The purple curve shows a stream with initial $r_0=15\,$kpc and $\sigma_v=0.6\,$km/s in each direction, evolved for 4.4 Gyr in a smooth potential.  The green curve shows the power spectrum for a stream with similar initial conditions, evolved for the same time in a FDM halo with $\lambda=0.6\,$kpc.}
    \label{fig:cluster_pk}
\end{figure}

Because the stream is no longer uniform in azimuthal angle $\phi$, but instead has some smooth profile $n(\phi)$, the stream profile now has a nonzero Fourier transform at $k>0$ even in the smooth potential.  
Therefore the smooth stream profile makes a contribution to the power spectrum, unlike the vanishing power spectrum of the uniform ring in the smooth potential.  This profile adds to the power spectrum of density fluctuations from FDM perturbations discussed in previous subsections. Fig.\ \ref{fig:cluster_pk} shows an example.  The purple curve shows the 1D density power spectrum for a stream originating from a compact cluster with $\sigma_v=3\times 10^{-3} v_c=0.6\,$km/s, evolved in a smooth potential for 4.4 Gyr.  On small scales, the power spectrum is consistent with shot noise, but at low $k$, the smooth profile of the stream dominates the power spectrum.  The shape of the purple curve is easy to understand from our discussion in \S\ref{sec:dispersion}.  Initially, when the cluster of test particles has zero spatial extent, the power spectrum is $P=1$.  Due to the velocity dispersion, this spectrum gets damped on small scales $k>k_d$, where the damping scale is $k_d=r_0/(\sigma_v t)$, which for $r_0=15\,$kpc, $\sigma_v=0.6\,$km/s, and $t=4.4\,$Gyr gives $k_d\approx 6$.  
This damping in Fourier space is simply the spreading of the particle positions in configuration space, i.e.\ the form of the damping is given by the overall (smooth) profile of the stream particles as a function of $\phi$.

The green curve in Fig.\ \ref{fig:cluster_pk} shows the corresponding FDM stream, which started from similar initial conditions and was evolved for 4.4 Gyr, in a halo with FDM $\lambda=0.6\,$kpc.
The FDM stream's power spectrum resembles that of the smooth stream at low $k$, because the streams in both cases have similar lengths.  But at high $k$, on scales much smaller than the stream length, the FDM power spectrum is larger than the smooth power spectrum by orders of magnitude, just as we found from the ring simulations.  The shape of the FDM power spectrum is similar to the results from the ring simulations, with $P\sim k^{-1}$ behavior which breaks to a steeper profile at a damping scale $k_d$ set by the initial velocity dispersion.  However, the location of the break seen in Fig.\ \ref{fig:cluster_pk} may seem surprising. We just argued that for these parameter values, we have $k_d\approx 6$, which agrees well with the smooth simulation (purple curve), but the break in the FDM power spectrum (green curve) occurs at a much smaller scale, $k_d \gtrsim 100$.

The reason for this behavior may be understood from Fig.\ \ref{fig:phiL}.  As the figure shows, the spread in angular momentum $L$ (and hence the angular frequency $\Omega$) is much smaller at a given location than the global dispersion across the entire stream.  This is a consequence of Liouville's theorem.  As the stream elongates in the $\phi$ direction over time, the local dispersion in $\Omega$ and $L$ across a fixed physical scale (like the de Broglie wavelength) must decrease accordingly, to conserve phase space density.  Since FDM perturbations are not all produced at time $t=0$, but instead are generated continuously over time, during which the local $v_\phi$ velocity dispersion steadily decreases, then it makes sense that the effective damping scale in the power spectrum would be considerably smaller (higher wavenumber) than the above estimate for $k_d$ using the global velocity dispersion.

\begin{figure*}
    \centering
    \includegraphics[width=0.48\textwidth]{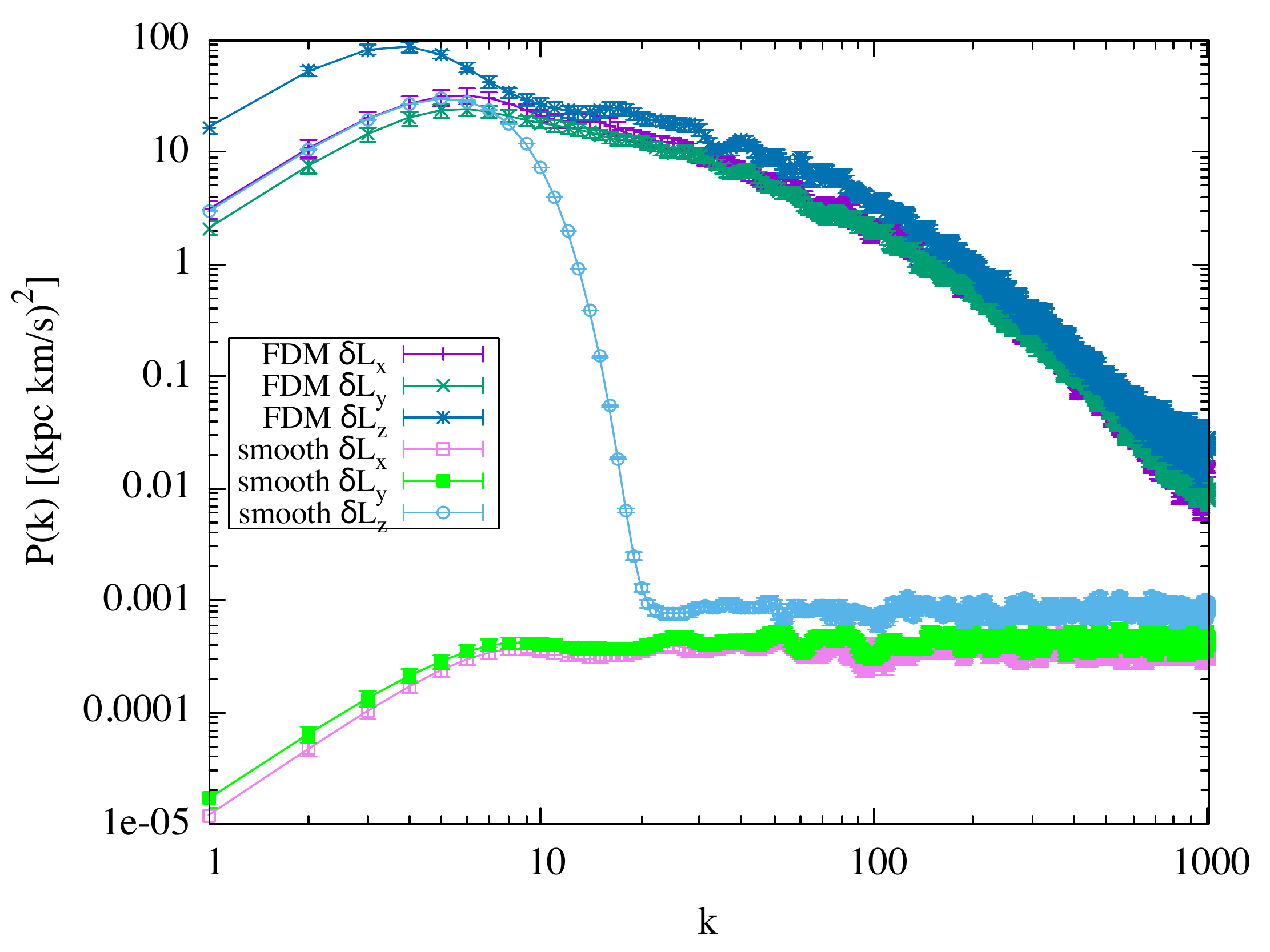}
    \includegraphics[width=0.48\textwidth]{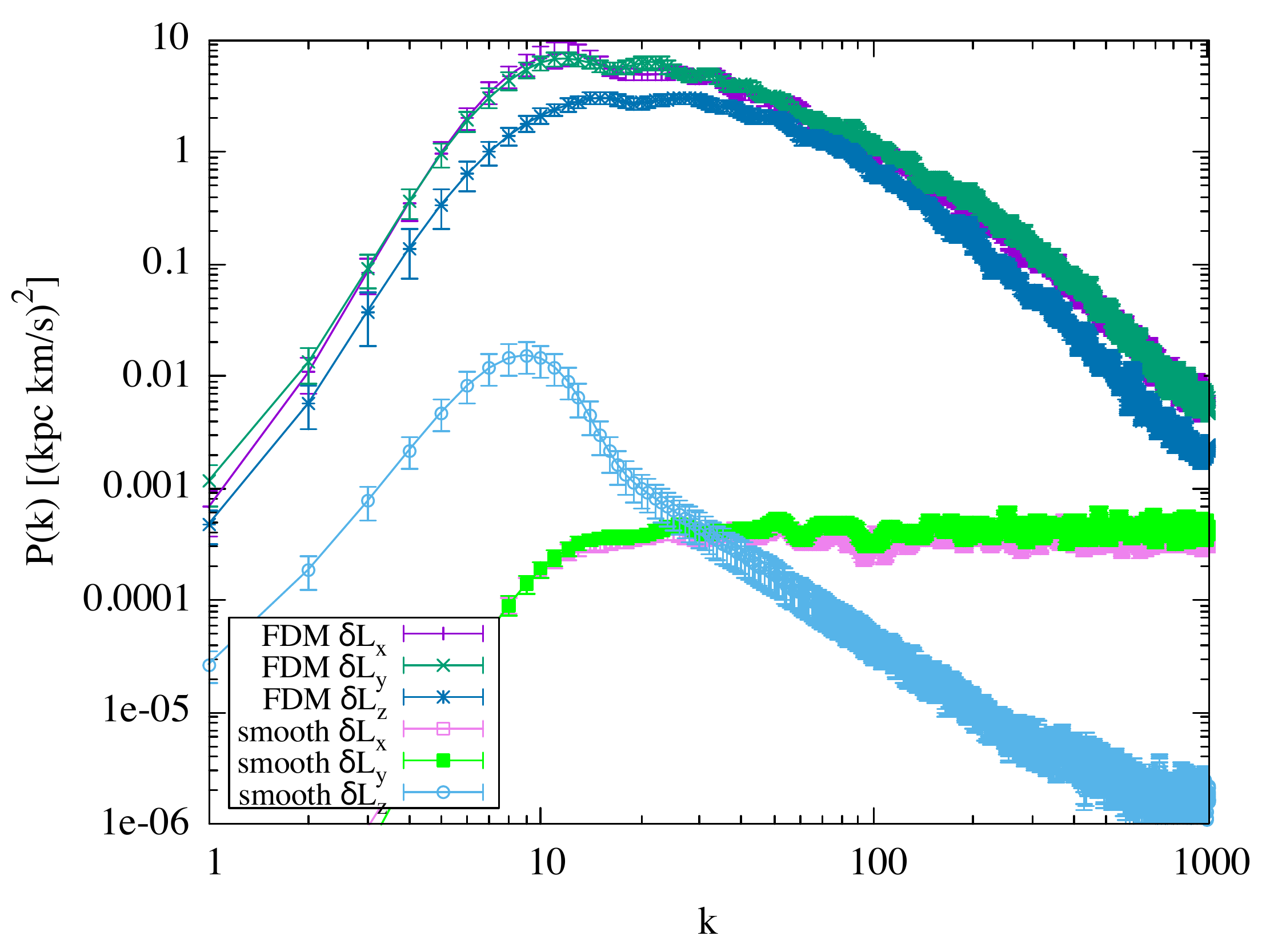}
    \caption{Angular momentum correlations.  Both panels plot the power spectrum of the product $(n(\phi)/{\bar n})\,\delta L(\phi)$ for streams in a smooth potential (lighter-colored curves) and a FDM potential with $\lambda=0.6\,$kpc (darker-colored curves).  In the left panel, we define $\delta L(\phi)=L(\phi)-\langle L \rangle$, and in the right panel, we use $\delta L(\phi)=L(\phi)-L_3(\phi)$, where $L_3$ is a cubic polynomial in $\phi$.}
    \label{fig:dLPk}
\end{figure*}

Another way to understand the same argument can be seen in Fig.\ \ref{fig:dLPk}, which plots the power spectrum of angular momentum perturbations.  One difficulty in computing the power spectrum of the specific angular momentum $L$ is that the number density $n(\phi)$ vanishes at locations where there are no particles, which makes the specific angular momentum $L(\phi)$ undefined beyond the ends of the stream.  For this reason, rather than measuring the two-point correlations of $L$, we will instead compute the correlations of the angular momentum density $nL$.  Because the mean $\langle L\rangle$ is much larger than the fluctuations $\delta L$, this would make the power spectrum of $nL$ merely a rescaled version of the density power spectrum.  Therefore, we compute the power spectrum of $n\,\delta L/{\bar n}$, where $\delta L = L - \langle L\rangle$, and the mean $\langle L\rangle$ is computed from all particles in the stream.  This is shown in the left panel of Fig.\ \ref{fig:dLPk}.  The lighter-colored curves are for the smooth simulation, and the darker-colored curves are for the FDM simulation.  As we saw in Figure \ref{fig:phiL}, even in the smooth potential there is a nearly linear relationship between $\phi$ and $L_z$, and this smooth $L(\phi)$ relation significantly affects the shape of the angular momentum power spectrum.  To remove this effect, in the right panel we plot the power spectrum of $n/{\bar n}$ times $\delta L=L-L_3(\phi)$, where $L_3$ is the cubic polynomial that best fits the mean relation between $L$ and $\phi$.  We subtract polynomial fits for all 3 components of $L$, even though this correction is most important for $L_z$.  Note that, in the right panel of Fig.\ \ref{fig:dLPk}, subtracting the best-fit cubic relation between $L$ and $\phi$ has reduced the shot noise contribution to the $L_z$ power spectrum by orders of magnitude at high $k$.  This reflects the behavior we noted in Fig.\ \ref{fig:phiL}, where the local dispersion in $L_z$ was much smaller than the global dispersion.  We do not see a similar reduction in the shot noise for $L_x$ or $L_y$, because those quantities are essentially uncorrelated with $\phi$.  If we were to redo Fig.\ \ref{fig:phiL} using $L_x$ or $L_y$ instead of $L_z$, we would see a similar global spread in the y-axis, but the scatter in $L$ would be almost uncorrelated with the scatter in $\phi$.

\subsubsection{Gradual escape}

\begin{figure}
    \centering
    \includegraphics[width=0.48\textwidth]{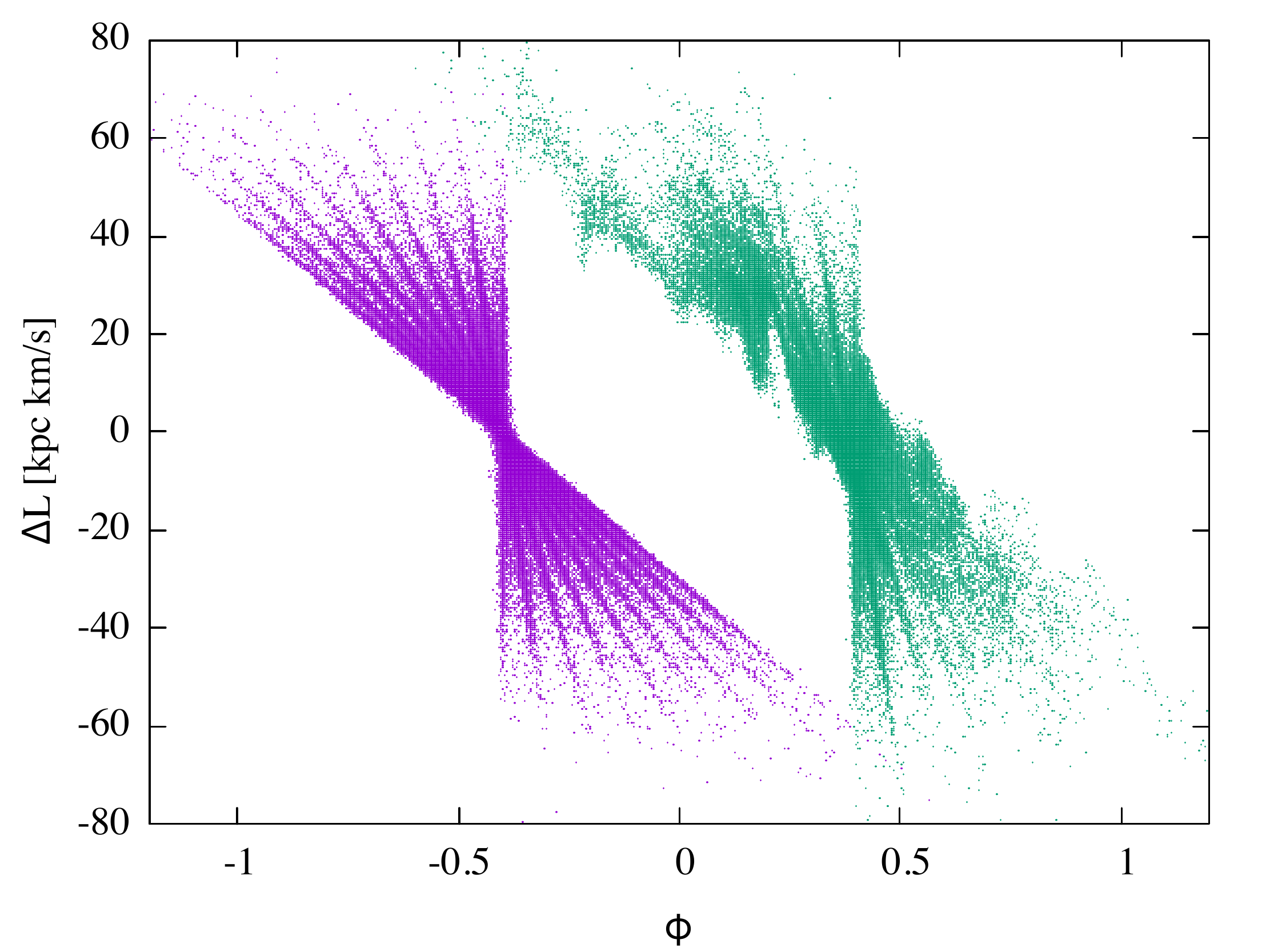}
    \caption{Similar to Fig.\ \ref{fig:phiL}, but for streams grown gradually over time.  Purple points were evolved in a smooth halo, and green points in a FDM halo.}
    \label{fig:phiL_gradual}
\end{figure}

\begin{figure}
    \centering
    \includegraphics[width=0.48\textwidth]{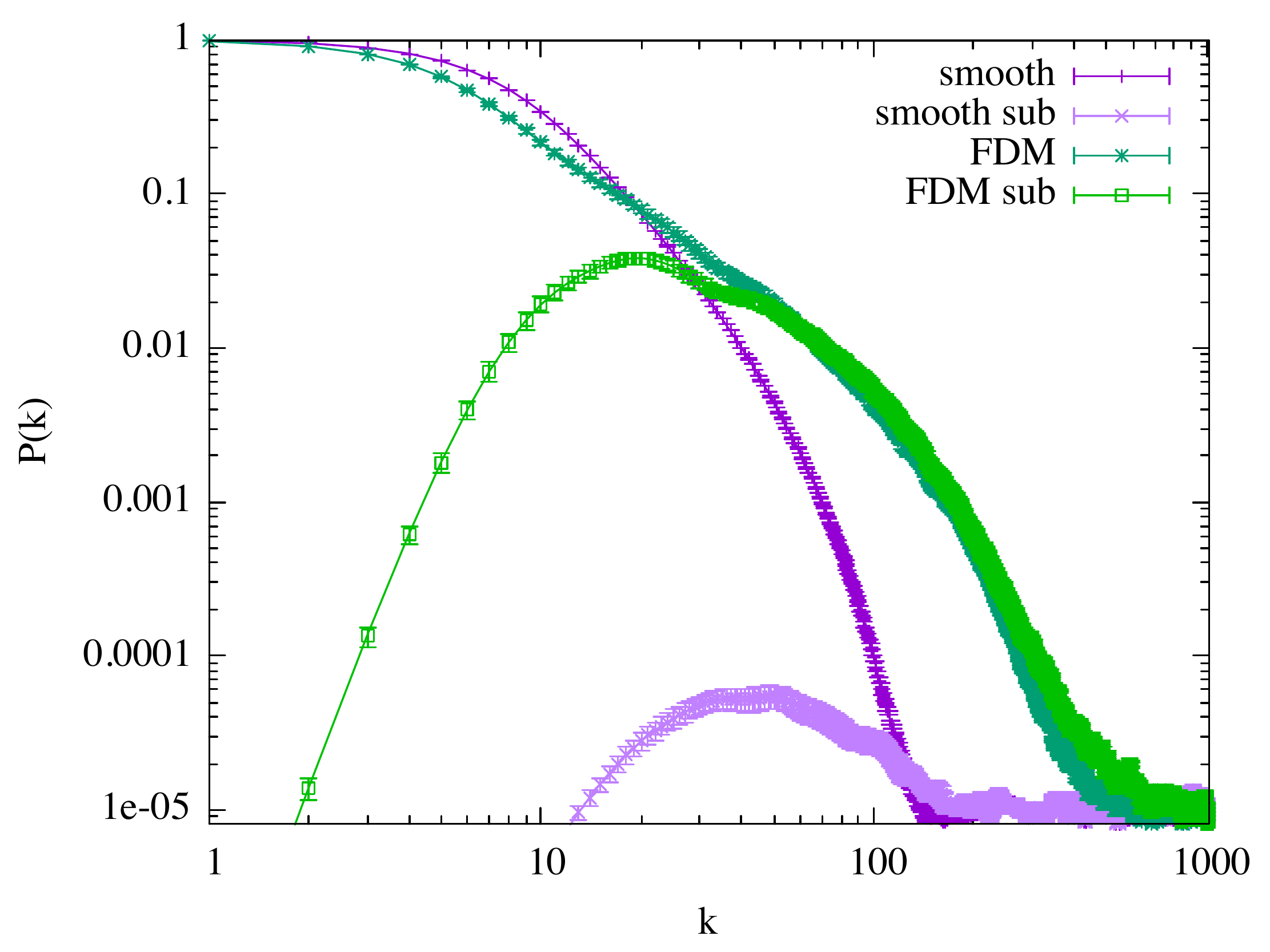}
    \caption{Density power spectra for streams in which particles are injected at a constant rate $dN/dt$, just like Fig.\ \ref{fig:phiL_gradual}.  The darker curves show power spectra for smooth potentials (dark purple) and FDM (dark green).  The lighter curves show power spectra when we subtract the best-fit density profile given by Eqn.\ \eqref{eqn:fit}.  This removes almost all of the structure in the smooth stream, but leaves unaffected the small-scale structure in the FDM stream.}
    \label{fig:Pk_gradual}
\end{figure}

The previous subsection showed results for simulations in which streams were created instantaneously. More realistically, streams grow gradually over time, as stars escape their clusters through a combination of tides and disk/bulge shocking near their pericenters.  Figures \ref{fig:phiL_gradual} and \ref{fig:Pk_gradual} show the effect this has on FDM perturbations.  We repeat our previous simulations, but instead of adding all particles to the stream at the start of the simulation, we instead add them at a constant rate $dN/dt$.  Fig.\ \ref{fig:phiL_gradual} shows the distribution of particles as a function of angle $\phi$ and angular momentum $L$, and the resulting plot looks similar to what we might have expected from combining the results from Fig.\ \ref{fig:phiL} over a range of times.  One possibly surprising new feature is that the distribution of points is not smooth in the $\phi-L$ plane, but instead the density of points is enhanced along discrete bands.  This effect, called ``epicyclic overdensities'' \cite{Kupper2010,Kupper2012,Amorisco2015}, is easy to understand for our simple simulations.  A particle's location $\phi$ is a function of its angular momentum $L$, the time it was injected into the stream $t_i$, and the time we observe the star, $t$.  For the simple case we consider here, in which the particle moves with the cluster at angular frequency $\Omega_0$ until time $t_i$, and then is placed on a new orbit with angular momentum $L$ after $t_i$, we have 
\begin{equation}
\phi = \Omega_0 t_i + \int_{t_i}^{t} \frac{L}{r^2(t)} dt.
\label{eqn:phase}
\end{equation}
Since orbits are not circular, then the instantaneous angular velocity ${\dot\phi}=L/r^2$ varies in time, oscillating about its average every radial period.  The distribution of angles $\phi$ for stars of a given $L$ and $t$ is 
\begin{equation}
\frac{dN}{d\phi} = \frac{dN/dt_i}{d\phi/dt_i},  
\label{eqn:dist}
\end{equation}
where $\phi$ is given by Eqn.\ \eqref{eqn:phase}, and the derivative in the right-hand side is evaluated at the $t_i$ that gives phase $\phi$ at time $t$. 
Even when the numerator (the injection rate) is constant in time, the denominator can vary significantly, since $\phi$ is not generally a monotonic function of injection time $t_i$.  When $d\phi/dt_i$ vanishes in the denominator of Eqn.\ \eqref{eqn:dist}, a caustic occurs, leading to a significant enhancement in $dN/(dL\,d\phi)$ shown in Fig.\ \ref{fig:phiL_gradual}.  This occurs both in smooth potentials and in FDM halos, but the density fluctuations in FDM halos eventually cause these oscillations to decohere, washing out the pileups seen at early times (compare green points to purple points).

Figure \ref{fig:Pk_gradual} shows the 1D density power spectra for streams that grow gradually over time.  The results are quite similar to Fig.\ \ref{fig:cluster_pk}.  One difference is that, even in the smooth halo, the power spectrum is significant down to degree scales ($l\sim 100$ instead of $l\sim 10$ in Fig.\ \ref{fig:cluster_pk}).  The reason is that the smooth profile of the stream peaks more sharply near the original cluster location, since many particles only recently entered the stream.  This more sharply peaked smooth profile has a power spectrum that extends to higher $k$, similar to the scales where we see FDM perturbations (compare darker purple and green curves in Fig.\ \ref{fig:Pk_gradual}).  Fortunately, it is straightforward to disentangle the power spectrum of the smooth profile from that of the FDM perturbations.  One method is simply to subtract the smooth profile before measuring the density power spectrum.  The lighter curves show the result of this subtraction.  In both cases, we fit the stream profile with a simple function, 
\begin{equation}
n_{\rm fit}(\phi) = \frac{A}{\sigma_2-\sigma_1}\left[ 
\Gamma\left(0,\frac{\phi^2}{2\sigma_2^2}\right) - \Gamma\left(0,\frac{\phi^2}{2\sigma_1^2}\right)\right],
\label{eqn:fit}
\end{equation}
where $\Gamma$ is the incomplete gamma function, and 
$A$, $\sigma_1$ and $\sigma_2$ are adjustable parameters.  This profile corresponds to a sum of Gaussians of dispersion varying linearly between $\sigma_1$ and $\sigma_2$.  For each stream, we determine the parameters by measuring the variance and kurtosis of the particles' angles $\phi$, and setting $\sigma_1$ and $\sigma_2$ to match those moments.  Subtracting this model profile from the measured density profile removes nearly all the power from the stream evolved in the smooth potential, while it cannot remove the small-scale power found in the FDM stream.  This is just one example to illustrate that the smooth profile is not degenerate with FDM perturbations, and quite likely more optimal methods can be constructed to disentangle the two contributions. 

Another potential source of structure in the stream is the rate $dN/dt_i$ at which stars enter the stream.  We have assumed constant injection rates corresponding to disruption by tides, but more generally $dN/dt_i$ will vary over time, for example when the cluster's orbit is highly eccentric, leading to significant variations in tidal stripping due to bulge and disk shocking.  A time-varying injection rate will produce structure qualitatively similar to that already seen in Fig.\ \ref{fig:phiL_gradual}, since (in the smooth halo) particles injected at a particular time roughly trace out diagonal lines in the $\phi-L$ or $\phi-\Omega$ planes.  We can therefore expect that time-varying injection rates cannot produce small-scale structure in the stream at locations distant from the star cluster, just as in Fig.\ \ref{fig:Pk_gradual}. The exception to this is when both $dN/dt_i$ and $dN/dL$ are not smooth, for example if $\langle|\Delta L|\rangle \gg \sigma_L$, i.e.\ if the scatter in the relative velocities between stars entering the stream is small compared to their velocities relative to the original cluster.  In general, though, we expect $\langle|\Delta L|\rangle \sim \sigma_L$, meaning that variations in the injection rate are not expected to produce larger effects in the power spectra than the caustics seen in Fig.\ \ref{fig:phiL_gradual}.

\section{Discussion}

In this paper, we have described a simple method to compute the effects of fuzzy dark matter on cold, thin, tidal streams in our Galaxy's halo.  
Our results indicate that FDM models can, in the parameter region of interest ($m\gtrsim 10^{-22}\,$eV), generate significant small-scale structure in tidal streams. Unlike in CDM, this small-scale structure results from interference fringes in the fluctuating DM density, rather than from bound subhalos, which are suppressed at $M \lesssim 10^{10}\,M_\odot$ in FDM for this FDM mass \cite{Schive16a}.
Any constraints on FDM models using stellar streams must quantitatively account for
perturbations from interference fringes in the FDM density; the effect of FDM cannot be captured by considering its suppressed subhalo mass function alone
\cite[e.g.][]{Schutz2020}.

We have argued that the power spectrum of structure along the stream provides a powerful probe of FDM fluctuations, potentially more discriminating than cruder measures like stream widths.  The width of a stream is sensitive not only to FDM perturbations but also to the initial velocity dispersion of stream stars, whereas fluctuations along the stream do not suffer from similar degeneracies.  Indeed, phase mixing associated with velocity dispersion tends to damp structure along streams, making the power spectrum a robust probe of halo substructure.

This raises the question of how well these power spectra may be measured.  In practice, constraints on FDM will depend on the details of the specific streams that are observed, but we can nonetheless make a rough estimate.  Potential sources of noise in measurements of the power spectrum include contamination from stars unassociated with the stream in question, Poisson fluctuations in the counts of stream stars, or depth variations that lead to spurious fluctuations in number density.  

Current observations of stellar streams have fore/background contamination of $N_{\rm stream}/N_{\rm bkgd} \approx 1$ (e.g., for the Pal 5 stream, $N_{\rm stream}/N_{\rm bkgd} \approx 0.5$ \cite{Ibata16a,Bovy2017}; for GD-1 with Gaia proper motion selection $N_{\rm stream}/N_{\rm bkgd} \approx 4$ \cite{Webb19a}, but for small $N_{\rm stream}$). 
Assuming these contaminating stars are unclustered, they
will reduce the signal-to-noise ratio in the overdensity power spectrum by a factor of $N_{\rm stream}/(N_{\rm stream}+N_{\rm bkgd})$.  The power spectra shown in Figures \ref{fig:cluster_pk} and \ref{fig:Pk_gradual} had no contamination, so the only noise source was Poisson fluctuations in the stream stars themselves, indicated by the white noise at a level $P\approx N_{\rm stream}^{-1}$ in the figures.  If we instead assume $N_{\rm bkgd}\approx N_{\rm stream}$, then in the overdensity power spectrum (i.e., the power spectrum of $\delta n/{\bar n}$) then the ratio between the signal power spectrum and the Poisson power spectrum decreases by about a factor of 2.  Additionally, these figures assume that $N_{\rm stream}=10^5$ stars are detected.
Current stream observations for GD-1 and Pal 5 have $N_{\rm stream} \approx 300$ to 3000. Therefore, if we instead observe $N_{\rm stream}=1000$ stars instead of $N_{\rm stream}=10^5$, while keeping $N_{\rm bkgd}\approx N_{\rm stream}$,
this would further reduce the SNR per mode by another factor of 100.   This would still give ${\cal O}(100)$ modes for which sample variance dominates over shot noise, giving a $\sim 7\sigma$
detection of the excess power spectrum. 
This is the SNR of the density power spectrum, and the SNR of other power spectra will depend on how well those quantities are measured.  For example, in Fig.\ \ref{fig:zz} we showed the power spectrum of vertical displacements from the mean orbital plane, and the SNR of that power spectrum should be similar to the density power spectrum, since the errors on the sky coordinates of stars are negligible.  On the other hand, the clustering of derived quantities like action-angle coordinates may be much noisier, since they require precise 6D coordinates for stream stars.

The SNR of the density power spectrum can be significantly improved if we can detect greater numbers of stream stars and better remove contamination.  Both of these should be possible with observations from the upcoming LSST survey at the Vera Rubin Observatory \cite{LSST}, which will survey approximately half the sky to depths exceeding those of existing large-area surveys like DES or Pan-STARRS.
Similarly, the Roman Space Telescope \cite{RomanST} will be able to observe stars in stellar streams down to the bottom of the stellar mass function \cite{Pearson19a}. The high photometric precision of these future instruments allows streams to be separated from the contamination through narrow filters in color--magnitude space, leading to $N_{\rm stream}/N_{\rm bkgd} \approx 2$ to 10 while simultaneously increasing $N_{\rm stream}$ to $\approx 10^4$ or higher. Then the SNR would only be reduced by a factor of $\lesssim 15$ compared to Fig.\ \ref{fig:Pk_gradual}.  This helps, not only by increasing the overall SNR of the power spectrum, but also by increasing the range of scales over which individual modes are detected above shot noise, allowing us to constrain the detailed shape of $P(k)$.
Such observations would allow us to clearly distinguish between the FDM density power spectrum and that of the smooth stream in Fig.\ \ref{fig:Pk_gradual}.  

In this paper, we have focused on perturbations arising from FDM interference fringes.  In addition to this form of substructure, halos  contain smaller subhalos as well, which also perturb tidal streams.  Both of these forms of substructure are present in FDM halos, and their relative contributions depend strongly on the FDM mass.  We can expect the subhalo abundance to be suppressed at least as strongly as the abundance of isolated halos is suppressed.  For $m\sim 10^{-22}\,$eV, the halo mass function is suppressed below $M_J \lesssim 10^{10} M_\odot$,  and for other FDM masses the suppression scale is expected to behave as $M_J \propto m^{-3/2}$ \cite{Schive16a,Kulkarni2020}.  Therefore, we can expect significant suppression of subhalos below mass $M \lesssim 10^9 - 10^{10} M_\odot$ for the FDM parameters we have considered.  Subhalos this massive are not present in large numbers inside of MW hosts, and produce individually distinctive effects on streams like gaps that can be identified and removed.

In CDM models, however, subhalos are present even at much lower masses.  It is quite straightforward to work out the stream perturbation power spectra produced by a population of subhalos, and in Appendix \ref{sec:subhalo} we sketch the argument.  Generally, the stream power spectra produced by CDM subhalos are quite similar in form to those produced by FDM perturbations, due to the universality of fold caustics.  A measurement of the density power spectrum of a tidal stream would therefore only tell us that the halo has substructure, but would not necessarily favor FDM substructure over CDM substructure.  One possible way to distinguish a CDM origin from a FDM origin would be to compare streams found at different Galactocentric radii $r$. As the appendices explain, in FDM the power spectrum scales as $(G{\bar\rho})^2\lambda^3$, whereas subhalos generate power spectra scaling as $(GM)^2 {\bar n}$, where $M$ is the subhalo mass and ${\bar n}$ is the subhalo number density near the stream.  Evidently, the power spectrum from FDM perturbations grows much more steeply with decreasing radius ($\propto {\bar\rho}^2$) than the power spectrum from subhalos ($\propto {\bar n}$).  Detection of streams over a range of Galactocentric radii could help to elucidate the nature of dark matter.

\acknowledgements
ND is supported by the Centre for the Universe at Perimeter Institute.
Research at Perimeter Institute is supported in part by the Government of Canada through the Department of Innovation, Science and Economic Development Canada and by the Province of Ontario through the Ministry of Colleges and Universities. This research was enabled in part by resources provided by Compute Ontario and Compute Canada. JB received financial support from NSERC (funding reference number RGPIN-2020-04712), an Ontario Early Researcher Award (ER16-12-061), and from the Canada Research Chair program.
LH is supported by a Simons Fellowship and the US Department of Energy DE-SC0011941.
XL is supported by the Natural Sciences and Engineering Research Council of Canada (NSERC), funding reference \#CITA 490888-16.
This work has made use of the GSL \cite{GSL}, FFTW \cite{FFTW}, and SHTns \cite{shtns} libraries, and we thank the respective authors for making their software publicly available.

\appendix

\section{Stream perturbations before caustics}
\label{sec:pre}

As discussed in \S\ref{sec:perts}, the two-point correlations of stream perturbations become dominated by the formation of fold caustics.  Prior to the development of folds, the stream perturbation power spectrum is simply related to the FDM density power spectrum.  To understand this behavior, recall (from Fig.\ \ref{fig:cl_rho}) that FDM perturbations have a nearly white-noise power spectrum with a form well approximated by Eqn.\ \eqref{eqn:whitenoise}.  These density perturbations lead to perturbations in the gravitational potential and acceleration, given by the Poisson equation.  If we insert Eqn.\ \eqref{eqn:whitenoise} into Eqn.\ \eqref{eqn:poisson}, noting that the cross-spectra $C_l(\Delta r)$ vanish for $\Delta r>\lambda$, then it is easy to see that at low $l$ the potential fluctuations have angular auto-spectra that scale as $\langle |\Phi_{lm}(r)|^2\rangle \propto (G{\bar\rho}(r))^2 \lambda^3 r / l^3$.  This is the 2D angular power spectrum; the corresponding 1D power spectrum of potential perturbations along a circle of radius $r$ therefore scales as $\langle |\Phi(k)|^2\rangle\propto (G{\bar\rho}(r))^2 \lambda^3 r / k^2$, where $k$ is the Fourier conjugate to azimuthal angle $\phi$. One way to see this is to note that both the 2D and 1D power spectra have the same Fourier transform, the  correlation function. Since the gravitational acceleration is $a=-\nabla\Phi$, then the 1D power spectrum of azimuthal acceleration perturbations scales as $\langle |a(k)|^2\rangle\propto(G{\bar\rho}(r))^2 \lambda^3/r$, independent of $k$, on large scales $k<2\pi r/\lambda$.  

\begin{figure}
    \centering
    \includegraphics[width=0.48\textwidth]{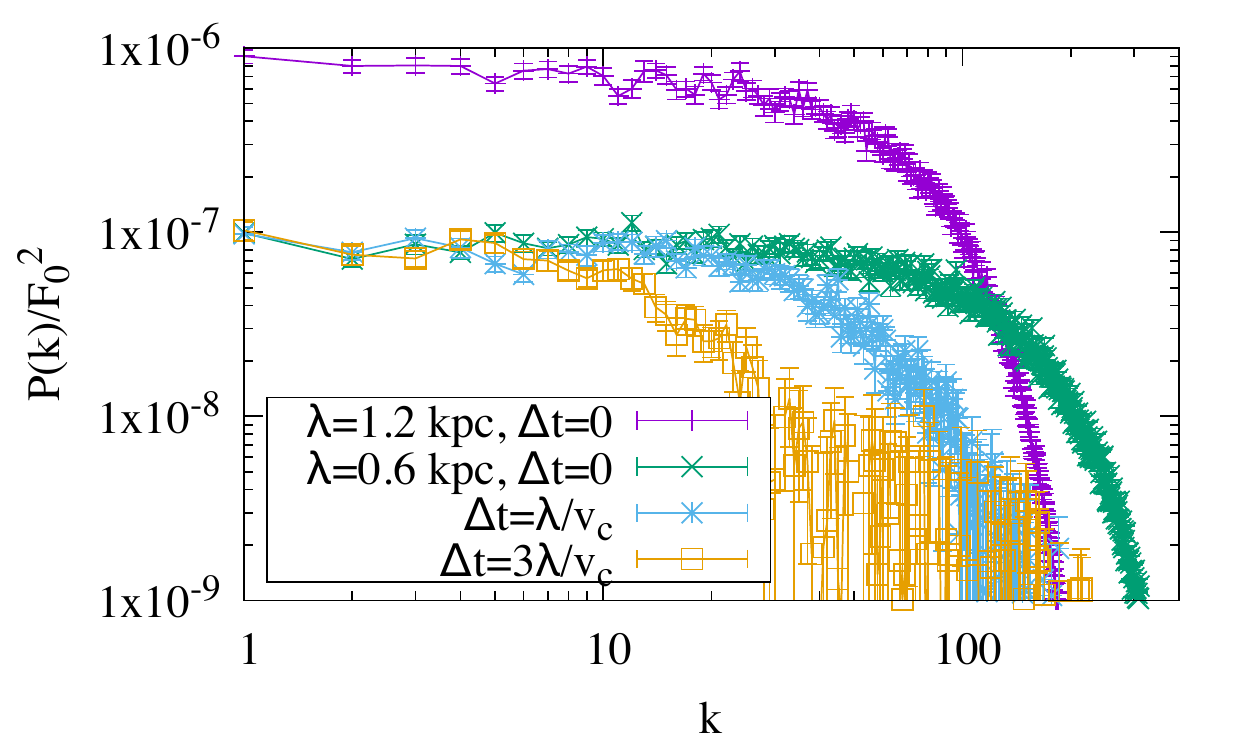}
    \caption{The different curves show the unequal time cross-correlations between the azimuthal acceleration fluctuations along a circular ring, as a function of FDM de Broglie wavelength $\lambda$ and the time interval $\Delta t$.  We normalize the acceleration by $F_0=v_c^2/r$.  As explained in the text, on scales larger than $2\pi r/\lambda$, the 1D power spectrum is nearly independent of $k$, and scales in amplitude like $\lambda^3$. Note that perturbations on scale $k$ lose coherence in time after interval $\Delta t\propto k^{-1}$, reflecting that fact that longer-wavelength perturbations remain coherent for longer times. }
    \label{fig:1dPk}
\end{figure}

A gravitational perturbation $a(k)$ will create velocity perturbations in the stream, according to ${\dot v}(k)=a(k)$, so $\Delta v(k)=a(k) \Delta t$, where $\Delta t$ is the coherence time of the perturbation mode $a(k)$.  The coherence time scales as $\Delta t \sim r/(v_c k)$, where $v_c$ is the FDM velocity -- modes on larger scales remain coherent for longer times. Figure \ref{fig:1dPk} shows an example, illustrating that the 1D acceleration power spectrum scales as $\lambda^3$, and that the perturbations are coherent over a time scale $\Delta t\propto k^{-1}$.  The velocity perturbation $v(k)$ therefore undergoes a random walk, of step size $\Delta v(k)=a(k) \Delta t$, and with $N=t/\Delta t$ steps, giving a total variance after time $t$ of $\langle |v(k)|^2\rangle = \langle |a(k)|^2\rangle \Delta t\, t \propto (G{\bar\rho}(r))^2 v_c^{-1} \lambda^3 t/k$.

\begin{figure}
    \centering
    \includegraphics[width=0.48\textwidth]{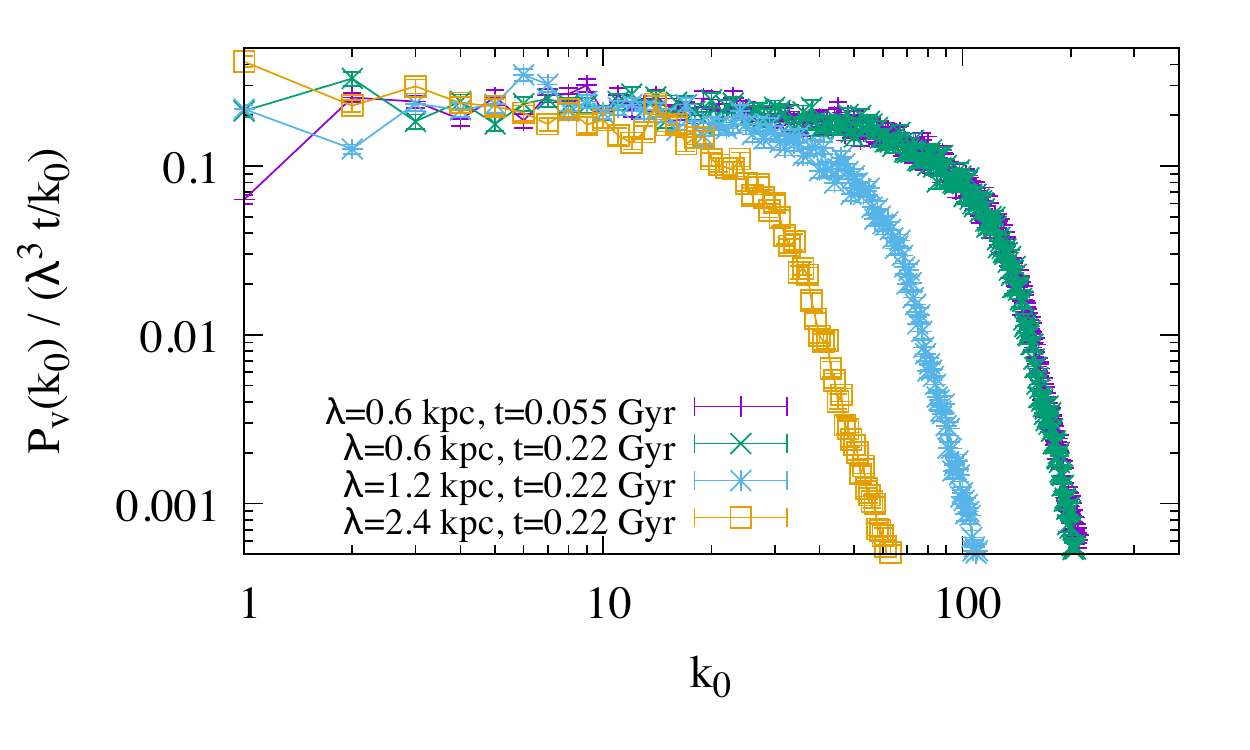}
    \caption{Power spectra of velocity perturbations.  We have divided the spectra by the expected scaling $P_v\propto \lambda^3 t/k$, to illustrate that the spectra do behave as expected at early times while perturbations remain small.}
    \label{fig:P_vv}
\end{figure}

Figure \ref{fig:P_vv} shows that we do indeed observe this behavior in the power spectrum of velocity fluctuations.  As in Fig.\ \ref{fig:variance}, we show spectra for particles initially spaced evenly in a ring of radius $r_0$, moving initially on circular orbits.  
To avoid complications due to folds, we determine $v$ as a function of each particle's initial (i.e., Lagrangian) coordinate $\phi_0$, and then Fourier transform along the $\phi_0$ direction to obtain $v(k_0)$ and the power spectrum $\langle v^*(k_0)v(k_0^\prime)\rangle=P_v(k_0)\delta_{k_0 k_0^\prime}$.  As expected, we find that that resulting velocity power spectra scale like $\lambda^3 t/k_0$, with exponential damping at $k_0>2\pi r_0/\lambda$.

These velocity perturbations give rise to displacement perturbations.  Since the displacement $\delta \phi = \int \delta v\,dt$, we expect that $\langle|\delta\phi(k)|^2\rangle\propto \langle |v(k)|^2\rangle t^2 \propto \lambda^3 t^3/k$.  The displacement perturbations give rise to density perturbations, since the overdensity fluctuation is $\delta n/{\bar n} = |1-d\delta\phi/d\phi_0|^{-1}-1$, where $\phi_0$ is the unperturbed coordinate.  Therefore, for small displacement perturbations, the density perturbations behave like $\langle|\delta n(k)|^2 \rangle \propto k^2 \langle|\delta\phi(k)|^2\rangle \propto \lambda^3 t^3 k$.  This explains the behavior seen at early times in Fig.\ \ref{fig:ring_pk}.

\begin{figure}
    \centering
    \includegraphics[width=0.48\textwidth]{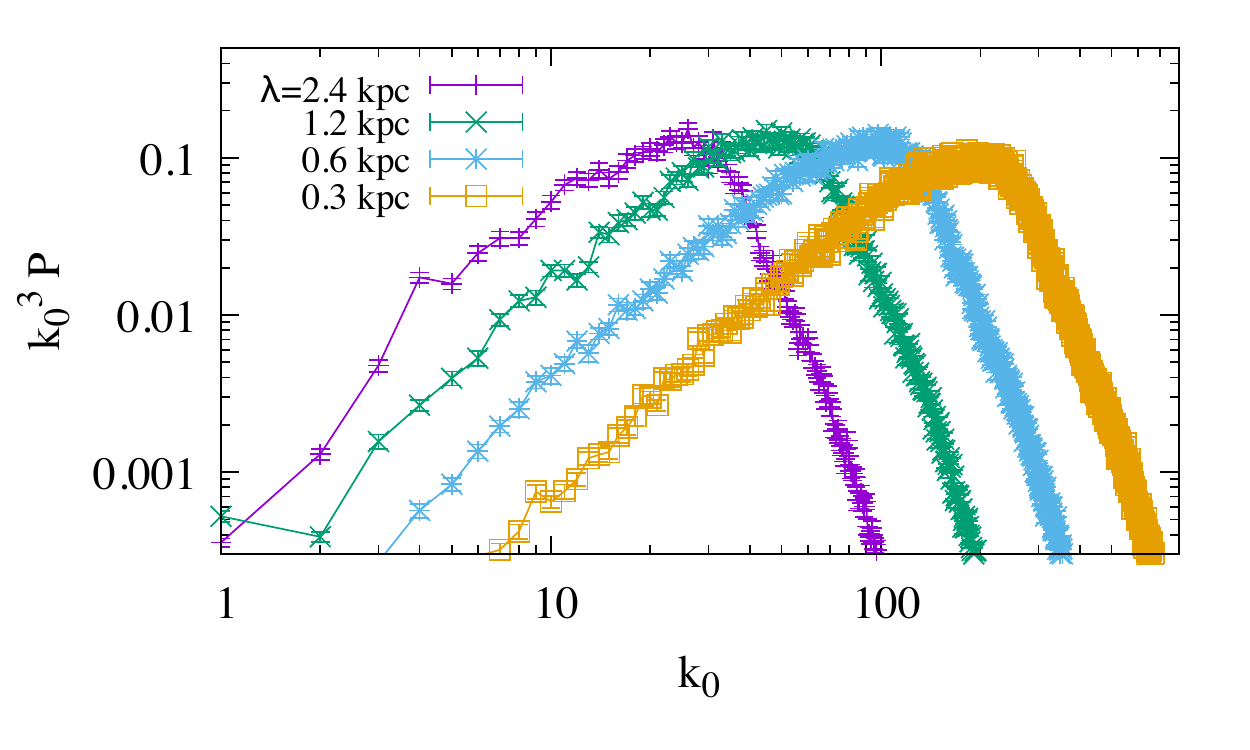}
    \caption{The various curves show $k_0^3 P_\phi$ at the time when fold caustics first develop in simulations with different $\lambda$.  As described in the text, the y-axis is essentially the variance in the divergence of the displacement field, and caustics occur when this variance becomes ${\cal O}(0.1)$.  The depicted times are $t=0.73\,$Gyr for $\lambda=0.6\,$kpc, and scale approximately as $t\propto \lambda^{-1/3}$ otherwise.
    }
    \label{fig:kphi}
\end{figure}

This behavior breaks down when folds begin to develop, causing $\delta n/{\bar n}$ to diverge at fold caustics.  Folds develop when regions occur where $|d\delta\phi/d\phi_0|\sim {\cal O}(1)$, so the time when caustics first appear should depend on the variance of $d\delta\phi/d\phi_0$, which is $\sum_{k_0} k_0^2 \langle|\delta\phi(k_0)|^2\rangle\sim k_{\rm max}^3 \langle|\delta\phi(k_{\rm max})|^2\rangle$, where $k_{\rm max} = 2\pi r_0/\lambda$.  As we noted above, the displacement power spectrum scales as $\langle|\delta\phi(k)|^2\rangle \propto t^3 \lambda^3/k$ for $k<k_{\rm max}$, so the variance of $d\delta\phi/d\phi_0$ grows as $t^3 \lambda^3 k_{\rm max}^2 \propto t^3 \lambda$.  We therefore expect that fold caustics will first develop at a time $t\propto \lambda^{-1/3}$.  This is indeed what we find.
In Fig.\ \ref{fig:kphi}, we plot $k_0^3 \langle|\delta\phi(k_0)|^2\rangle$, where $k_0$ is the Fourier conjugate of the unperturbed (Lagrangian) coordinate $\phi_0$.  As the figure shows, fold caustics first develop when the total variance $k_0^3 \langle|\delta\phi|^2\rangle\approx 0.1$, nearly independent of $\lambda$, and the time when this occurs scales as $t\propto \lambda^{-1/3}$.  For our fiducial case of a stream at $r_0=15\,$kpc with $m=10^{-22}$eV, this time is $t\sim 0.7\,$Gyr.  

\section{Stream perturbations from subhalos}
\label{sec:subhalo}

Consider a population of subhalos, each of mass $M$ and size $R$, with mean number density profile within the host halo given by ${\bar n}(r)$.  If the host has mass density profile ${\bar\rho}(r)$, then the mass fraction in subhalos of mass $M$ is $f = M{\bar n}/{\bar\rho}$. Let us assume that the subhalos are unclustered within their host halo, which is a good description for host halos of mass similar to the MW \cite{Chamberlain2015}.  Then the angular power spectrum of density fluctuations at radius $r$, associated with this population of subhalos, takes the form 
\begin{equation}
C_l(r) = f^2 {\bar\rho}^2 W^2(l\theta) N^{-1},
\label{eqn:subhaloCl}
\end{equation}
where $\theta\propto R/r$ is an effective angular size for the subhalos at radius $r$, $W$ is related to the subhalos' internal density profile, and $N=4\pi r^2 R_{\rm eff} {\bar n}$ is the average number of subhalos found at radius $r$, for some effective size $R_{\rm eff} \sim R$ that depends on the subhalos' internal profiles.  In the language of the halo model \cite{Seljak2000}, this is the 1-subhalo (or Poisson) contribution to the power spectrum, and we neglect the 2-subhalo contribution, under the assumption that subhalo clustering is negligible in hosts like the MW, similar to the calculations of the lensing substructure power spectrum  \cite{Hezaveh2016}.  

Eqn.\ \eqref{eqn:subhaloCl} is the auto-spectrum, and similar to FDM fluctuations, we expect substructure density fluctuations to become uncorrelated for annuli separated by $\Delta r > R_{\rm eff}$.  Given the density power spectrum, we can repeat the argument of Appendix \ref{sec:pre} to estimate the power spectrum of stream perturbations.  The 2D angular power spectrum of potential fluctuations at low $l$ is $\langle |\Phi_{lm}(r)|^2\rangle \propto (GM)^2 {\bar n} r / l^3$, implying that the 1D power spectrum along a circle of radius $r$ behaves as $\langle |\Phi(k)|^2\rangle\propto (GM)^2 {\bar n} r / k^2$.  Therefore $\langle |a(k)|^2\rangle\propto(GM)^2 {\bar n}/r$, and $\langle |v(k)|^2\rangle \propto (GM)^2 {\bar n} t\, v_c^{-1} k^{-1}$, for $k \ll r/R$.  Just as for FDM perturbations, the displacement power spectrum scales as $\langle |\delta\phi(k)|^2\rangle \propto t^2 \langle |v(k)|^2\rangle$, and the density power spectrum will scale as $k^2 \langle |\delta\phi(k)|^2$, until caustics develop.

This is the stream power generated by subhalos of mass $M$.  For subhalos of a range of mass, we simply add the power spectra from each mass bin, $P_\phi \propto \int M^2 d{\bar n}/d(\log M) d(\log M)$.  Since $d{\bar n}/d\log M \propto M^{-0.9}$ \cite{Diemand2008}, then in the linear regime of stream perturbations, the stream power spectrum on large scales ($k\ll\theta^{-1}(M_{\rm max})$) is strongly dominated by perturbations from the most massive subhalos \cite{Bovy2017}. At linear order, small subhalos  dominate the small scale power spectrum at $k\gg\theta^{-1}(M_{\rm max})$, assuming that the subhalo internal profile $W(l\theta)$ declines sufficiently steeply at $l\theta\gg1$.  However, because the linear power spectrum grows in time $\propto t^3$, nonlinearity will eventually develop.
Once perturbations become nonlinear and caustics arise, the most massive subhalos dominate the 1D power spectrum on all scales.  Depending on the age of a given stream, we could observe it during the linear regime or the nonlinear regime, which means that any constraints derived from actual streams in our Galaxy will require careful comparison with numerical simulations.

\section{The stream density power spectrum}
\label{sec:ZA}

In this Appendix, we derive an expression for the stream density power spectrum that covers both the pre-caustic and caustic regimes. Consider for simplicity the set-up in \S \ref{sec:powerspectra}, \ref{sec:dispersion} and \ref{sec:others}: a circular ring of particles undergoing gravitational scattering by substructure. The trajectory of a particle along the stream is described by:
\begin{eqnarray}
\phi = \chi + \Omega t + \Delta\phi (\chi, t) \, ,
\end{eqnarray}
where $\chi$ is the initial position, $\phi$ is the position at time $t$, $\Omega t$ is the displacement if there were no substructure, and $\Delta\phi$ is the displacement due to scattering by substructure.
In our set-up, all particles along the stream start suffering scattering at the same initial time $t=0$. 

The stream density at time $t$ is
\begin{eqnarray}
n(\phi) = \int d\chi \, \delta_D (\phi - \chi - \Omega_0 t - \Delta \phi(\chi, t)) \, ,
\end{eqnarray}
where we have set the initial density to unity.
The Fourier transform of $n$ is
\begin{eqnarray}
n(k) = \int d\chi \, e^{-ik(\chi + \Omega_0t + \Delta \phi(\chi, t))}
\, .
\end{eqnarray}
The power spectrum $P_n (k)$ at time $t$ is defined by:
\begin{eqnarray}
&& \langle n(k) n^* (k') \rangle = 2\pi \delta_D (k-k') P_n (k) \, \nonumber \\
&& \quad \quad \quad \quad \quad = \int d\chi d\chi' e^{-ik\chi + ik'\chi'} e^{-i(k-k')\Omega_0 t} \nonumber \\
&& \quad \quad \quad \quad \quad \quad \quad \langle e^{-ik \Delta\phi(\chi,t) + ik' \Delta\phi(\chi',t)} \rangle \, ,
\end{eqnarray}
where the ensemble average is over realizations of $\Delta\phi$. We could have subtracted the mean from $n$ before squaring it to obtain the power spectrum, in which case $P_n (k)$ would vanish at $k=0$. As long as $k \ne 0$, doing so or not makes no difference. 
Translational invariance implies the expectation value depends only on $\chi - \chi'$. 
Changing the variable of integration from $\chi$ and $\chi'$ to $\bar\chi \equiv (\chi + \chi')/2$ and $\delta \chi \equiv \chi - \chi'$, we can integrate over $\bar\chi$ and obtain:
\begin{eqnarray}
P_n (k) = \int d\chi \, e^{-ik \chi} \, \langle e^{-ik (\Delta\phi(\chi, t) - \Delta\phi(0, t))} \rangle \, ,
\end{eqnarray}
where we have replaced $\delta\chi \rightarrow \chi$.
This derivation follows that of Taylor and Hamilton \cite{Taylor:1996ne} for the power spectrum in the Zeldovich approximation.
If $\Delta \phi$ is a Gaussian random field, the above expectation value can be computed exactly
giving
\begin{eqnarray}
\label{Pnxi}
P_n (k) = \int d\chi \, e^{-ik \chi} e^{-k^2 \xi (\chi, t)} \, ,
\end{eqnarray}
where 
\begin{eqnarray}
&& \xi (\chi, t) \equiv {1\over 2} \langle (\Delta\phi(\chi, t) - \Delta\phi(0, t))^2 \rangle
\nonumber \\
&& \quad \quad = \langle \Delta\phi(0, t) {}^2 \rangle - 
\langle \Delta\phi(\chi, t) \Delta\phi(0, t) \rangle \, .
\end{eqnarray}
It is worth noting that $\Delta \phi$, which is related to the gravitational potential
$\Phi$, is probably not Gaussian random. If the halo is composed of waves with random phases,
the wavefunction is Gaussian random, but the density and the potential are not \cite{Hui:2020hbq}.
Nonetheless, the Gaussian approximation appears to give a $P_n$ that is in reasonable agreement with numerical computations.

Eqn. (\ref{Pnxi}) is general: it holds regardless of whether caustics are important or not. If $k^2 \xi$ is small (a sufficient condition is if $\langle k^2 \langle \phi(0, t)^2 \rangle$ is small),
we can expand out the exponent to obtain the power spectrum in the perturbative, pre-caustic regime:
\begin{equation}
\label{Pnperturb}
P_n (k) \sim \int d\chi e^{-ik\chi} (1 - k^2 \xi(\chi, t)) \sim k^2 P_{\Delta\phi} (k) \, ,
\end{equation}
where we have dropped delta function terms which can be ignored if $k \ne 0$,
and $P_{\Delta\phi}$ is the angular displacement power spectrum at time $t$:
$P_{\Delta\phi} (k) = \int d\chi e^{-ik\chi} \langle \Delta\phi(\chi, t) \Delta\phi(0, t) \rangle$. 

If $k^2 \langle \Delta \phi(0,t)^2 \rangle > 1$, we are in the caustic regime. In this case,
$\chi$ must be small to keep $\xi$ small. By definition, $\xi(0, t) = 0$. Taylor expanding it in $\chi$, we see that the first order term vanishes by parity, and so $\xi(\chi, t) \propto \chi^2$ for small $\chi$. 
The structure of the integral in 
Eq. (\ref{Pnxi}) is thus roughly $P_n (k) \sim \int d\chi e^{-ik\chi} e^{-k^2\chi^2...}$. The $k$ dependence is thus $P_n (k) \sim 1/k$, which agrees with the estimate obtained by considering the caustic profile (square of the Fourier transform of $1/\sqrt{\phi}$ gives $1/k$). 
One can estimate the time of caustic formation by checking when $k^2 [k P_{\Delta \phi} (k)/2\pi]_{\rm max}$ exceeds unity, where $[\,\,]_{\rm max}$ is evaluated at the scale that maximizes it.

To estimate $P_{\rm \Delta\phi}$, one can use
\begin{eqnarray}
\label{app:Deltaphi3}
&& \Delta \phi (\chi, t) \sim {1\over r_0^2} \int_0^t dt' \int_0^{t'} dt'' \partial_\phi \Phi(\phi, t'') \Big|_{\phi = \chi + \Omega t''} \nonumber \\
&& \quad \quad = {1\over r_0^2} \int_0^t dt'' (t - t'') \partial_\phi \Phi(\phi, t'') \Big|_{\phi = \chi + \Omega t''} \, ,
\end{eqnarray}
which follows from including contributions from both angular momentum perturbation and radial perturbation (non-oscillatory part thereof). Gradient of $\Phi$ in orthogonal directions in principle contribute, but can be shown to be subdominant in the large $t$ limit. Squaring the above, one obtains the displacement correlation function:
\begin{eqnarray}
&& \langle \Delta\phi(\chi, t) \Delta\phi(0, t) \rangle \sim \nonumber \\
&&  {t^3 \over 3 r_0^4} \int dt'' \langle \partial_\phi \Phi (\phi, 0) \Big|_{\phi = \chi} \partial_\phi \Phi (\phi, t'') \Big|_{\phi = \Omega t''} \rangle  \, 
\end{eqnarray}
where the Limber approximation has been made, and $t''$ can be thought of as being integrated over infinity. 
The integral involves an unequal time correlation function. See Appendix \ref{sec:pre} on how to estimate it.

\newcommand{\jcap}{J.\ Cosm.\ Astroparticle Phys.}
\newcommand{\mnras}{Mon.\ Not.\ Roy.\ Astron.\ Soc.}
\newcommand{\aj}{Astron.\ J.}
\newcommand{\apjl}{Astrophys.\ J.\ Lett.}
\newcommand{\aap}{Astron.\ Astrophys.}

\bibliography{fdm}

\end{document}